\documentclass[
reprint,
 amsmath,amssymb,
 aps,prd
]{revtex4-1}

\usepackage{graphicx}
\usepackage{dcolumn}
\usepackage{bm}
\usepackage{float}
\usepackage{subfigure}
\usepackage{natbib}
\usepackage{silence}
\begin{document}

\preprint{APS/123-QED}

\title{Angular asymmetries as a probe  \\ 
for anomalous contributions to $HZZ$ vertex at the LHC}

\author{N.~Belyaev}
\affiliation{National Research Nuclear University Ò(Moscow Engineering Physics Institute)Ó, 31 Kashirskoe Shosse, Moscow 115409, Russia}%

\author{R.~Konoplich}
\affiliation{Department of Physics, New York University, 4 Washington Place, New York, NY 10003, USA}
\altaffiliation[Also at ]{Physics Department, Manhattan College, 4513 Manhattan College Parkway, Riverdale, New York, NY 10471, USA.} 

\author{L.~Egholm Pedersen}
\affiliation{%
 Niels Bohr Institute, University of Copenhagen, Blegdamsvej 17 Kobenhavn, Denmark
}%

\author{ K.~Prokofiev}
\affiliation{Department of Physics, Hong Kong University of Science and Technology, Clear Water Bay, Kowloon, Hong Kong}
\altaffiliation[Also at ]{Institute for Advanced Study, HKUST, Clear Water Bay, Kowloon, Hong Kong.}



\date{\today}

\begin{abstract}

In this article, the prospects for studying the tensor structure of the $HZZ$ vertex 
 with the LHC experiments are presented. 
 The  structure   of  tensor couplings  in Higgs di-boson decays is investigated 
 by measuring the  asymmetries and by studing the shapes of the final state angular distributions. 
 The expected background  contributions, detector resolution, and trigger and selection efficiencies 
 are taken into account. The potential of the LHC experiments to discover sizeable non-Standard Model 
 contributions to the $HZZ$ vertex with $300\;{\rm fb}^{-1}$  and $3000\;{\rm fb}^{-1}$  
 is demonstrated.

\end{abstract}

\maketitle

\setlength{\textfloatsep}{10pt plus 1.0pt minus 2.0pt}

\setlength{\intextsep}{10pt plus 1.0pt minus 2.0pt}

\label{sec:intro}
\section{Introduction}
In the Summer of 2012, the CMS and ATLAS Collaborations at the LHC  reported the 
discovery of a new neutral resonance in searches for the Standard Model
Higgs boson.  This discovery was later confirmed 
by analyses of the full LHC Run-I dataset by both collaborations \cite{cms_higgs, atlas_higgs}.
It was demonstrated that the new particle with a  mass around $125.5$~GeV was dominantly produced 
via the gluon-fusion process and decays into pairs of gauge bosons: $\gamma \gamma$, $ZZ$
and $WW$. The observed production and decay modes identified the discovered particle as 
a neutral boson.  The subsequent measurement of its couplings to fermions and 
bosons demonstrated  the compatibility of the discovered resonance with the expectations
for the Standard Model Higgs boson within available statistics \cite{atlas_couplings, Khachatryan:2014kca, Chatrchyan:2012jja}. 

In the Standard Model, electroweak symmetry breaking via the Higgs mechanism requires the 
presence of a single neutral Higgs boson with spin 0 and even CP-parity. Theories beyond 
the Standard Model often require an extended Higgs sector featuring several neutral Higgs bosons 
of both even and odd CP-parity. In such a case, mixing between Higgs boson  CP-eigenstates 
is possible. The Higgs boson mass eigenstates observed in experiment  may thus have  mixed 
CP-parity. Such an extension of the Higgs sector is important because effects of CP violation in the SM
are too small and, in particular, cannot explain the baryon asymmetry of the Universe. 

%
Dedicated studies of spin and parity of the Higgs candidate discovered by ATLAS and 
CMS showed that its dominant spin and parity are compatible with $J^{CP}=0^{++}$ \cite{atlas_spin, Chatrchyan:2012jja, Khachatryan:2014kca}. 
The dataset of about $25\;{\rm fb}^{-1}$ currently collected by each of the major LHC experiments allows to set an 
upper limit on the possible CP-odd contribution. The sensitivity is expected to improve with larger datasets to be collected at the LHC.  

There have been many works on direct measurement of CP violation in the Higgs sector ~\cite{Chang:1993jy,Grzadkowski:1995rx,Gunion:1996vv,Grzadkowski:1999ye,Grzadkowski:2000hm,Plehn:2001nj, Choi:2002jk, Buszello:2002uu,Hankele:2006ma,Godbole:2007cn,Keung:2008ve,Berge:2008dr,Cao:2009ah,DeRujula:2010ys,Gao:2010qx,Berge:2011ij,Bishara:2013vya,Harnik:2013aja,Berge:2013jra,Modak:2013sb,Chen:2014gka, Gainer:2013rxa, Avery:2012um, Gainer:2014hha, Chen:2013waa, Stolarski:2012ps}. In this paper the sensitivity of LHC experiments to observe  CP-mixing effects with $300\;{\rm fb}^{-1}$ and $3000\;{\rm fb}^{-1}$  is evaluated using the method of angular asymmetries.  

This paper is organised as follows. In Section II
observables sensitive to
CP violation in the $HZZ$ vertex are discussed. The spin-0 model, a Monte Carlo production of 
signal and background, and a lagrangian parametrisation for CP-mixing measurements are discussed 
in Section III.
In Section IV 
the expected sensitivity of the LHC experiments 
to the CP-violation effects based on angular asymmetries is presented. Constraits are set on the
contribution of anomalous couplings to the $HZZ$ vertex. Section V
introduces the measurement technique based on observables fit. Exclusion regions for the mixing angle are
presented. Section VI
gives the overall summary of obtained results.

\label{sec:theory}
\section{Observables}
In this paper we study the sensitivity of final state observables to the CP 
violating $HZZ$ vertex in the process:
\begin{equation}
gg \to H \to ZZ \to 4l.
\label{eq:ggH}
\end{equation}

Following the notation introduced in \cite{Gao:2010qx}, the general scattering amplitude describing
 interactions of a spin-zero boson with the gauge bosons is given by:
\begin{align}
A(X \to VV) = \frac{1}{v} \big(g_{ 1} m_{V}^2 \epsilon_1^{\ast} \epsilon_2^{\ast} + g_{ 2} f_{\mu \nu}^{\ast (1)}f^{\ast (2)\mu \nu}  \nonumber \\
 +\; g_{ 4}  f^{\ast(1)}_{\mu \nu} {\tilde f}^{\ast (2)\mu \nu}\big)\,.
\label{eq:ampl} 
\end{align}
Here the  $f^{(i){\mu \nu}} = \epsilon_i^{\mu}q_i^{\nu} - \epsilon_{i}^\nu q_i^{\mu} $ 
is the field strength tensor of a gauge boson 
with momentum $q_i$ and polarisation vector $\epsilon_i$;
${\tilde f}^{(i)\mu \nu} = 1/2 \epsilon^{\mu \nu \alpha \beta} f_{\alpha \beta}$  
is the  conjugate field strength tensor.  The symbols $v$  and $m_{V}$ denote
the SM vacuum expectation value of the Higgs field and the mass of the gauge boson
respectively.
 
In the Standard Model, the only non-vanishing coupling of the Higgs  to $ZZ$ or $WW$ boson pairs
at tree-level is $g_1=2i$, while $g_{ 2}$   is generated through radiative corrections. 
For final states with at least one massless gauge boson, such as $\gamma \gamma$, $gg$ or 
$Z\gamma$, the SM interactions with the Higgs boson are loop-induced. 
These interactions are described by the coupling $g_2$. The coupling $g_4$ is associated
with the interaction of CP-odd Higgs boson with a pair of gauge bosons.
The simultaneous presence of CP-even terms $g_1$ and/or $g_2$ and the CP-odd term $g_4$ leads
to CP violation.  

In general, $g_i$ couplings can be complex and momentum dependent. However imaginary parts of
these couplings are generated by absorptive parts of the corresponding diagrams and 
expected to be small: approximately less than $1\%$. We further assume that the energy scale of 
new physics is around $1$~TeV or higher, so that the momentum dependence of the couplings can be neglected.
Thus, in the following we will consider $g_i$ couplings as real and momentum-independent.    

These assumptions are entirely consistent with the framework of an effective field theory (EFT) of the SM. If the energy scale of the new physics is much higher than the electroweak scale new effects can be described by an EFT with the SM Lagrangian supplemented by higher dimension operators of $d=6$. Such an approach was worked out in detail in \cite{Buchmuller:1985jz, Grzadkowski:2010es}

\begin{figure}[htbp]                                                                                                         
\centering
\includegraphics[width=1.0\linewidth]{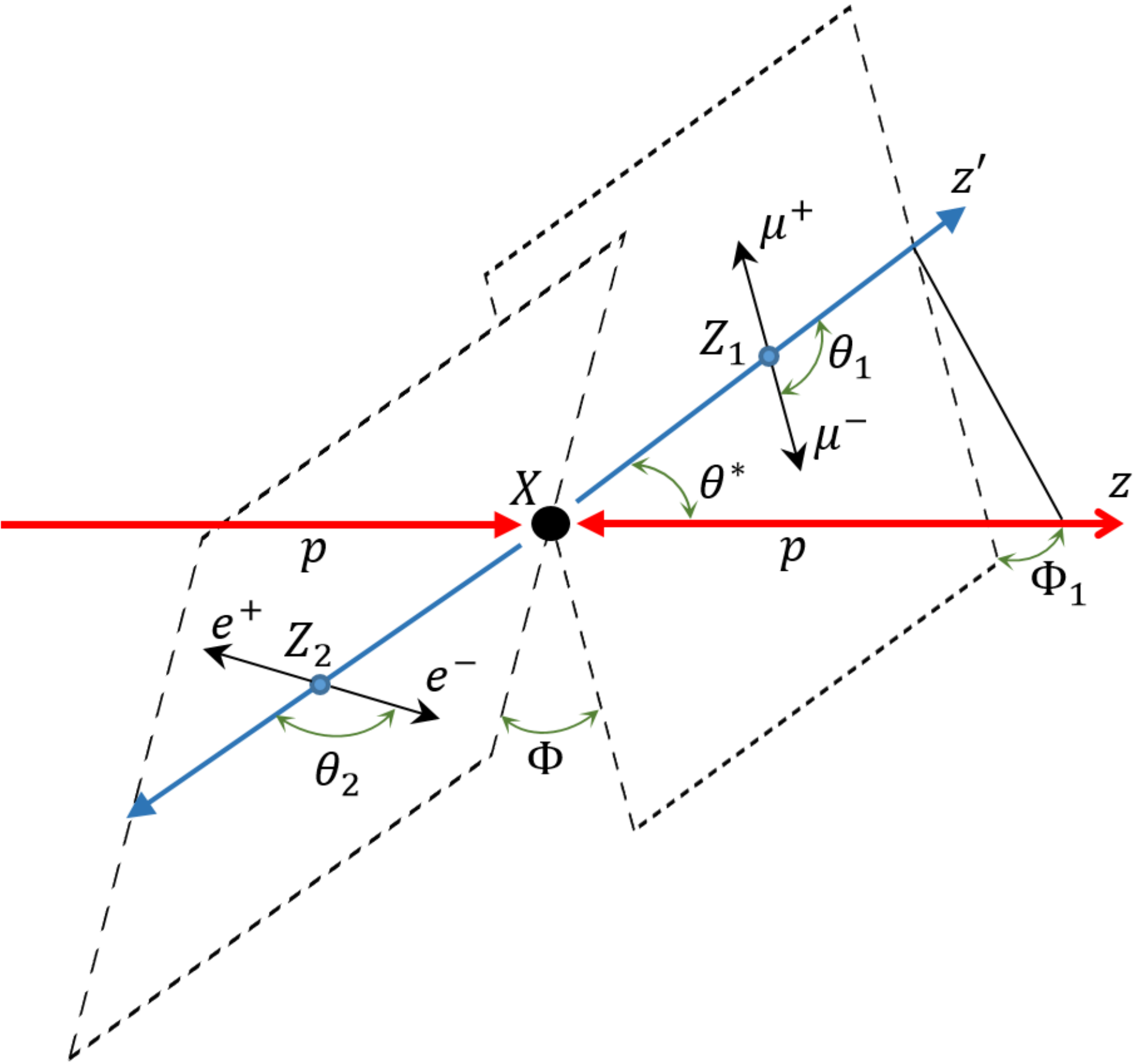}
\caption{Definitions of the CP-sensitive final state angular observables  in 
$gg\to H\to ZZ \to 4l$ decay.}

\label{fig:angles}         
\end{figure}

One of possibilities to study CP violation in the process of Eq. \ref{eq:ggH} is to analyse the shapes
of angular and mass distributions of the final state \cite{JHU2, JHU3}. The common choice of 
angular observables for this type of analysis is show in Fig.~\ref{fig:angles}.
   
A complimentary  approach is 
based on  studies of  angular-function asymmetries arising in the case of
CP violation. There are six observable functions proposed in \cite{asymmetries}. 
The first angular observable function is defined as follows:
\begin{equation}
O_1 = \frac{(\vec{p}_{2Z} - \vec{p}_{1Z}) \cdot
(\vec{p}_{3H} + \vec{p}_{4H})}{|\vec{p}_{2Z} - \vec{p}_{1Z}|
|\vec{p}_{3H} + \vec{p}_{4H}|} \;.   \nonumber
\end{equation}
Here $\vec{p}_i$, $i=1, \ldots 4$ are the 3-momenta of the final state leptons
in the order $l_1 \bar l_1 l_2 \bar l_2$.
The subscripts $Z$ and $H$ denote that the corresponding 3-vector is 
taken in the $Z$  or in the Higgs boson rest frames.   Using these definitions,  the second observable  function reads:
\begin{equation}
O_2 = \frac{(\vec{p}_{2Z} - \vec{p}_{1Z}) \cdot 
(\vec{p}_{4H} \times \vec{p}_{3H})}{|\vec{p}_{2Z} - \vec{p}_{1Z}| 
|\vec{p}_{4H} \times \vec{p}_{3H}|} \;.   \nonumber
\end{equation}
The third observable function $O_3$ is constructed  using $O_1$:
\begin{equation}
O_3 = O_1 \, O_{3a} \, O_{3b} \;,  \nonumber 
\end{equation}
where
\begin{equation}
O_{3a} = \frac{(\vec{p}_{4Z} - \vec{p}_{3Z}) \cdot 
(\vec{p}_{1H} \times \vec{p}_{2H})}{|\vec{p}_{4Z} - \vec{p}_{3Z}| 
|\vec{p}_{1H} \times \vec{p}_{2H}|} \;   \nonumber
\end{equation}
and
\begin{equation}
O_{3b} = \frac{(\vec{p}_{3Z} - \vec{p}_{4Z}) \cdot 
(\vec{p}_{1H} + \vec{p}_{2H})}{|\vec{p}_{3Z} - \vec{p}_{4Z}| 
|\vec{p}_{1H} + \vec{p}_{2H}|} \;.    \nonumber
\end{equation}
The remaining three observable functions are given by:
\begin{equation}
O_4 = \frac{[(\vec{p}_{3H}\times\vec{p}_{4H})\cdot \vec{p}_{1H}]
[(\vec{p}_{3H}\times\vec{p}_{4H})\cdot (\vec{p}_{1H}\times\vec{p}_{2H})]}{|\vec{p}_{3H}+\vec{p}_{4H}|^2|\vec{p}_{1H}+\vec{p}_{2H}||\vec{p}_{3Z}-\vec{p}_{4Z}|^2|\vec{p}_{1Z}-\vec{p}_{2Z}|^2/16} \;,   \nonumber
\end{equation}
\begin{equation} 
O_5 =
\frac{[(\vec{p}_{4H}\times\vec{p}_{3H})\cdot\vec{p}_{1H}][(\vec{p}_{1Z}
-\vec{p}_{2Z})\cdot\vec{p}_{3Z}]}{|\vec{p}_{3H}+\vec{p}_{4H}||\vec{p}_{3Z}-\vec{p}_{4Z}|^2
|\vec{p}_{1Z}-\vec{p}_{2Z}|^2/8} \;,   \nonumber
\end{equation}
and
\begin{equation}
O_6 = \frac{[(\vec{p}_{1Z}-\vec{p}_{2Z})\cdot (\vec{p}_{3H}+\vec{p}_{4H})]
[(\vec{p}_{3H}\times\vec{p}_{4H})\cdot \vec{p}_{1H}]}{|\vec{p}_{1Z}-\vec{p}_{2Z}|^2 |\vec{p}_{3H}+\vec{p}_{4H}|^2 |\vec{p}_{3Z}-\vec{p}_{4Z}|/4} \;.  \nonumber
\end{equation}
These observables are  related to the final state  angular variables defined in \cite{JHU2} and illustrated in Fig.~\ref{fig:angles}.
For instance, a trivial calculation yeilds:  $O_1 = \cos \theta_1$ and $O_2 = -\sin\phi \sin\theta_1$.
  
Note that the total cross section is CP even
(no interference between CP-even and CP-odd terms) and cannot be used to
detect the presence of CP violating terms in the $HZZ$ vertex.

\label{sec:mc}
\section{Spin-0 model and Monte Carlo production}

The dominant Higgs boson production mechanism at the LHC is gluon-fusion.
To simulate the production of a Higgs-like boson and its consequent decay into 
$ZZ$ and $4l$,  the MadGraph5 Monte Carlo generator \cite{Alwall:2011uj} was used. 
This generator implements the Higgs Characterisation model \cite{Artoisenet:2013puc}. 
The corresponding  effective Lagrangian describing the  interaction of the spin-$0$ Higgs-like 
boson with vector bosons  is given by:
\begin{align}
 {\cal L}_0^V =\bigg\{&
  c_{\alpha}\kappa_{\rm SM}\big[\frac{1}{2}g_{HZZ}\, Z_\mu Z^\mu 
                                +g_{HWW}\, W^+_\mu W^{-\mu}\big] \nonumber \\
  &\mkern -40mu -\frac{1}{4}\big[c_{\alpha}\kappa_{H\gamma\gamma}
 g_{H\gamma\gamma} \, A_{\mu\nu}A^{\mu\nu}
        +s_{\alpha}\kappa_{A\gamma\gamma}g_{A\gamma\gamma}\,
 A_{\mu\nu}\widetilde A^{\mu\nu}
 \big] \nonumber \\
  &\mkern -40mu -\frac{1}{2}\big[c_{\alpha}\kappa_{HZ\gamma}g_{HZ\gamma} \, 
 Z_{\mu\nu}A^{\mu\nu}
        +s_{\alpha}\kappa_{AZ\gamma}g_{AZ\gamma}\,Z_{\mu\nu}\widetilde A^{\mu\nu} \big] \nonumber \\
  &\mkern -40mu -\frac{1}{4}\big[c_{\alpha}\kappa_{Hgg}g_{Hgg} \, G_{\mu\nu}^aG^{a,\mu\nu} 
        +s_{\alpha}\kappa_{Agg}g_{Agg}\,G_{\mu\nu}^a\widetilde G^{a,\mu\nu} \big] \nonumber \\
  &\mkern -40mu -\frac{1}{4}\frac{1}{\Lambda}\big[c_{\alpha}\kappa_{HZZ} \, Z_{\mu\nu}Z^{\mu\nu}
        +s_{\alpha}\kappa_{AZZ}\,Z_{\mu\nu}\widetilde Z^{\mu\nu} \big] \nonumber \\
  &\mkern -40mu -\frac{1}{2}\frac{1}{\Lambda}\big[c_{\alpha}\kappa_{HWW} \, W^+_{\mu\nu}W^{-\mu\nu}
        +s_{\alpha}\kappa_{AWW}\,W^+_{\mu\nu}\widetilde W^{-\mu\nu}\big] \nonumber \\ 
  &\mkern -40mu -\frac{1}{\Lambda}c_{\alpha} 
    \big[ \kappa_{H\partial\gamma} \, Z_{\nu}\partial_{\mu}A^{\mu\nu}
         +\kappa_{H\partial Z} \, Z_{\nu}\partial_{\mu}Z^{\mu\nu}  \nonumber \\
  &\mkern -40mu +\big(\kappa_{H\partial W} \, W_{\nu}^+\partial_{\mu}W^{-\mu\nu}+h.c.\big)
 \big]
 \bigg\} X\,,
 \label{eq:lagrange}
\end{align}
 where $\Lambda$ is the new physics energy scale and the field strength tensors are defined as follows:
\begin{align}
 V_{\mu\nu} &=\partial_{\mu}V_{\nu}-\partial_{\nu}V_{\mu}\quad (V=A,Z,W^{\pm})\,, \nonumber \\
 G_{\mu\nu}^a &=\partial_{\mu}^{}G_{\nu}^a-\partial_{\nu}^{}G_{\mu}^a    
  +g_sf^{abc}G_{\mu}^bG_{\nu}^c\,.  \nonumber
\end{align}
The dual tensor $ \widetilde V_{\mu\nu}$ is defined as:
\begin{align}
 \widetilde V_{\mu\nu} =\frac{1}{2}\epsilon_{\mu\nu\rho\sigma}V^{\rho\sigma}.\,  \nonumber
\end{align}
The mixing angle $\alpha$ allows the production and decay of CP-mixed states and implies CP 
violation when $\alpha \neq 0$ or $\alpha \neq \pi/2$.
The definitions of effective tensor couplings  $g_{XVV'}$ are shown in Table~\ref{tab:gXVV}. 
\begin{table}
  \begin{center}
    \begin{tabular}{|r|c|c|c|c|}
      \hline
                        & $ZZ/WW$     & $\gamma\gamma$                                  & $Z\gamma$                              & $gg$                     \\      
      \hline
      $v \cdot g_{HVV'}$ & $2m_{Z/W}^2$ & $ \frac{\strut 47\alpha_{\rm EM}}{\strut 18\pi}$ & $ C \frac{94 \cos^2\theta_W-13}{9\pi}$ & $-\frac{\alpha_s}{3\pi}$ \\
      $v \cdot g_{AVV'}$ &  0          & $\frac{\strut 4\alpha_{\rm EM}}{\strut 3\pi}$    & $2 C \frac{8\cos^2\theta_W-5}{3\pi}$   & $\frac{\alpha_s}{2\pi}$  \\ 

      \hline
    \end{tabular}
    \caption{Definitions of effective tensor couplings $g_{XVV'}$ introduced in Eq.~(\ref{eq:lagrange}) in 
      units of  the Higgs vacuum expectation $v$. The symbol $C$ is defined as: $C=\sqrt{\frac{\alpha_{\rm EM}G_F m_Z^2}{8\sqrt{2}\pi}}$.}
    \label{tab:gXVV}
  \end{center}
\end{table}

The Lagrangian in Eq.~(\ref{eq:lagrange}) is an effective Lagrangian with $U(1)_{EM}$ symmetry.  
It parametrizes all possible Lorentz structures, is not $SU(2)\times U(1)$ invariant and does not assume that 
the Higgs boson belongs to a doublet of the weak $SU(2)$ group. Interaction terms corresponding to 
a Lagrangian of this type do not necessarily  form a complete basis. However, this form is convenient 
for analysis of   experimental data, as it relates in a simple way effective couplings and quantities 
observed in experiments. Note that there is a different and very popular EFT approach ~\cite{Grzadkowski:2010es} 
to studies of the Higgs boson sector based on a complete  set of operators of dimension six (the so-called Warsaw basis).

The relations between parameters of the Lagrangian of Eq.~(\ref{eq:lagrange}) and tensor couplings of the 
effective amplitude of Eq.~(\ref{eq:ampl}) can be derived from Feynman rules. The corresponding conversion 
coefficients  are shown in Table~\ref{tab:sigbcgsu3}.

\begin{table}[htbp]
  \begin{center}
    \begin{tabular}{|c|c|c|c|c|c|}
      \hline
      Coupling & ZZ & WW & $\gamma \gamma$ & Z $\gamma$ & $gg$  \\
      \hline\hline
      & & & & &   \\      
      $g_1$/$2ic_a$ & $k_{SM}$ & $k_{SM}$ & -  & -  & -  \\
      & & & & & \\
      $g_2$/$2ic_a$ & $\tilde{K}_{HZZ}$ & $\tilde{K}_{HWW}$ & $\tilde{K}_{H \gamma \gamma}$  & $\tilde{K}_{HZ \gamma}$ & $\tilde{K}_{Hgg}$ \\
      & & & & & \\
      $g_4$/$2is_a$ & $\tilde{K}_{AZZ}$ & $\tilde{K}_{AWW}$ & $\tilde{K}_{A \gamma \gamma}$  & $\tilde{K}_{AZ \gamma}$ & $\tilde{K}_{Agg}$ \\
      & & & & & \\
      $g^{''}_1$/$2ic_a$ & $\tilde{K}_{H\partial Z}$ & $Re(\tilde{K}_{H\partial W})$ & -  & -  & -  \\
      & & & & & \\
      $g^{'''}_1$/$2ic_a$ & -  & $iIm(\tilde{K}_{H\partial W})$ & -  & -  & -  \\
      & & & & & \\
      \hline
    \end{tabular}
    \caption{Conversion coefficients  between  parameters of the Lagrangian of Eq.~(\ref{eq:lagrange}) and tensor couplings of the 
      effective amplitude of Eq.~(\ref{eq:ampl}).}\label{tab:sigbcgsu3}
  \end{center}
\end{table}  

In this table the following definitions are used:
\begin{equation}
\tilde{K}_{XVV'}=\frac{1}{4}\frac{v}{\Lambda}\tilde{g}_{XVV'}k_{XVV'} , \nonumber
\end{equation}
\begin{equation}
\tilde{K}_{H\partial V}=\frac{1}{2}\frac{v}{\Lambda}\left(\frac{\Lambda_{1}}{m_{V}}\right)^{2}k_{H\partial V} ,  \nonumber
\end{equation}
\begin{equation}
c_{\alpha}=\cos\alpha,  \; {\rm and} \;  s_{\alpha}=\sin\alpha. \nonumber
\end{equation}
Here $X$ denotes either $H$ or $A$ and the index $VV'$ denotes the final 
state gauge boson pair. The effective couplings $\tilde{g}_{XVV'}$ are defined 
as follows:
\begin{itemize}
 \item{ In the case of $ZZ$ or $WW$ interactions,  $\tilde{g}_{XVV'} = 1$;}
 \item{ For $\gamma\gamma$,  $Z\gamma$ and  $gg$ interactions, 
couplings $\tilde{g}_{XVV'}$ are equivalent to the couplings   $g_{XVV'}$
defined in Table~\ref{tab:gXVV}.}
\end{itemize}
The couplings $\tilde{K}_{H\partial V}$, where $V=W,Z,\gamma$, correspond to the so-called contact terms of 
the Higgs Characterisation Lagrangian of Eq.~(\ref{eq:lagrange}).  These contact terms can be reproduced in 
the amplitude of Eq.~(\ref{eq:ampl}) by re-parametrising the $g_1$ coupling in the following form \cite{JHU_manual}:
\begin{equation}
g_{1}\left(q^{2}_{1},q^{2}_{2}\right)=g_{1}^{SM}+g^{'2}_{1}\frac{\left|q_{1}^{2}\right|+\left|q_{2}^{2}\right|}{\Lambda_{1}^{2}}+g^{'3}_{1}\frac{\left|q_{1}^{2}\right|-\left|q_{2}^{2}\right|}{\Lambda_{1}^{2}}.  \nonumber
\end{equation}
This equation represents the leading terms of the form factor expansion.
 In the case of complex   $k_{H\partial W}$, the momenta of the $W$ bosons should be 
 assigned as follows: $q_1$ for $W^-$ and $q_2$ for $W^+$. 
 In the case of $HZ\gamma$ interaction with a real photon, the term proportional to $k_{H\partial \gamma}$
 vanishes.

In the following we will consider a model based on the Lagrangian of Eq.~(\ref{eq:lagrange}) in which the mixing is provided 
by the simultaneous presence of the Standard model CP-even term and a non-Standard model CP-odd term
in the $HZZ$ decay vertex. The signal Monte Carlo samples used in this analysis are produced using 
the  Higgs Characterisation model parameters presented in Table~\ref{model}. 
\begin{table}[htbp]
  \centering
  \begin{tabular}{|c|c|c|c|c|c|}
    \hline
    $k_{SM}$ & $k_{HZZ}$ & $k_{AZZ}$ & $k_{Hgg}$ & $k_{Agg}$ & $\Lambda, GeV$  \\
    \hline
    $1$ & $0$ & $28.6$ & $1$ & $1$ & $10^{3}$  \\
    \hline
  \end{tabular}
  \caption{\label{model} Parameters of Higgs Characterisation model used for Monte-Carlo simulation of signal samples.}
\end{table}\\

The coefficient $k_{AZZ}$ was chosen such that it provided equal cross sections for  decays of CP-odd and CP-even 
Higgs states:  $\sigma(c_{\alpha}=0)=\sigma(c_{\alpha}=1)$.  The tensor couplings for the decay vertex 
corresponding to the amplitude of Eq.~(\ref{eq:ampl}) can be restored using the following relations: $g_2=2ic_\alpha$ and $g_4=2is_\alpha\tilde{K}_{AZZ}$,
where $\tilde{K}_{AZZ}=1.76$.
It is noted that the factor $2i$ is not important in the study of asymmetries because it defines the overall
cross-section normalisation.

The signal samples were produced using the MadGraph5 Monte Carlo generator \cite{Alwall:2011uj}. 
These samples were created in the range of mixing angles $-1\leq\cos\alpha\leq1$ in steps of $0.05$.
The dominant background processes $q\bar{q}\to ZZ, Z\gamma$ were also simulated with MadGraph5.

After simulation of signal and background events at $\sqrt{s} = 14$~TeV, the parton showering was performed 
using the PYTHIA6 Monte Carlo generator \cite{pythia}. Generic detector effects were included by using the 
PGS package~\cite{Alwall:2011uj}. 
The main detector parameters used for this simulation are presented in Table~\ref{tab:det_tune}.
\begin{table}[htbp]
  \centering
  \begin{tabular}{l l}
    \hline
    \hline
     Parameter & Value\\
    \hline   
    Electromagnetic calorimeter resolution  $\cdot\sqrt{E}$ & 0.1 \\ 
    Hadronic calolrimeter resolution$\cdot\sqrt{E}$ & 0.8 \\ 
    MET resolution & 0.2 \\
    Outer radius of tracker (m) & 1.0 \\
    Magnetic field (T) & 2.0 \\
    Track finding efficiency & 0.98 \\
    Tracking $\eta$ coverage & 2.5 \\
    $e/\gamma$ $\eta$ coverage & 2.8 \\
    Muon $\eta$ coverage & 2.8 \\
    \hline
    \hline
    
  \end{tabular} 
  \caption{\label{tab:det_tune} 
Tuning parameters used to simulate detector effects with PGS package.}
 \end{table}   
For comparison, the expected acceptance, efficiencies and resolutions of the ATLAS and CMS detectors 
 of the LHC can be found in ~\cite{ATLAS_paper, CMS_paper}. \\
Finally a kinematic selection was applied. 
It was required that candidates decayed to two same flavour oppositely charged lepton pairs. 
If several of such candidates could be reconstructed in an event, the leptons pairs with invariant masses closest
to the on-shell Z mass where chosen. Each individual lepton had a psudorapidity 
$|\eta| < 2.5$ and transverse momentum $p_{T} > 7~$GeV. The most energetic lepton should satisfy $p_{T} > 20~$GeV whereas the 
second (third) similarly had $p_{T} > 15~$GeV ($p_{T} > 10~$GeV).
The invariant mass of the on-shell Z boson was in the mass window $(50,106)~$GeV while the off-shell Z boson $m_{Z*}>20~$GeV. Only 
Higgs candidates in the signal region $115~$GeV$ < m_{H} < 130~$GeV where considered. 
The selection is a simplified version of the one presented in ~\cite{atlas_higgs}.

\label{sec:asym}
\section{\label{sec:asym} Asymmetries}
For each observable $O_i$ sensitive to CP violation, the corresponding asymmetry can be 
defined as:
\begin{equation}
A_i = \frac {N(O_i>0) - N(O_i < 0)} {N(O_i>0) + N(O_i < 0)},
\label{eq:asym}
\end{equation}
where $N$ is the number of  events with the observable less or greater than zero. 
Integrating the corresponding decay probabilities, it can be shown that these asymmetries 
directly probe the tensor couplings defined in the amplitude of Eq.~(\ref{eq:ampl}) \cite{asymmetries}. 
The value of $A_1$ is proportional to $Im(g_4)$, while $A_2, A_3, A_4, A_5$
and $A_6$ probe the values of $Re(g_4)$ and $Im(g_2)$ respectively.

Analysis of asymmetries sensitive to CP-violation for the process of Eq.~(\ref{eq:ggH}) was performed in \cite{asymmetries}.
In this section we extend this analysis by including effects of parton showering,
hadronization, generic detector effects and contributions from the 
irreducible  $q\bar{q}\to ZZ/Z\gamma \to 4l$ background.
Lepton interference in the final state and the contribution of two 
off-shell $Z$-bosons are also taken into account.

\begin{figure}[htbp]
\begin{minipage}[h]{0.49\linewidth}
\center{\includegraphics[width=1\linewidth]{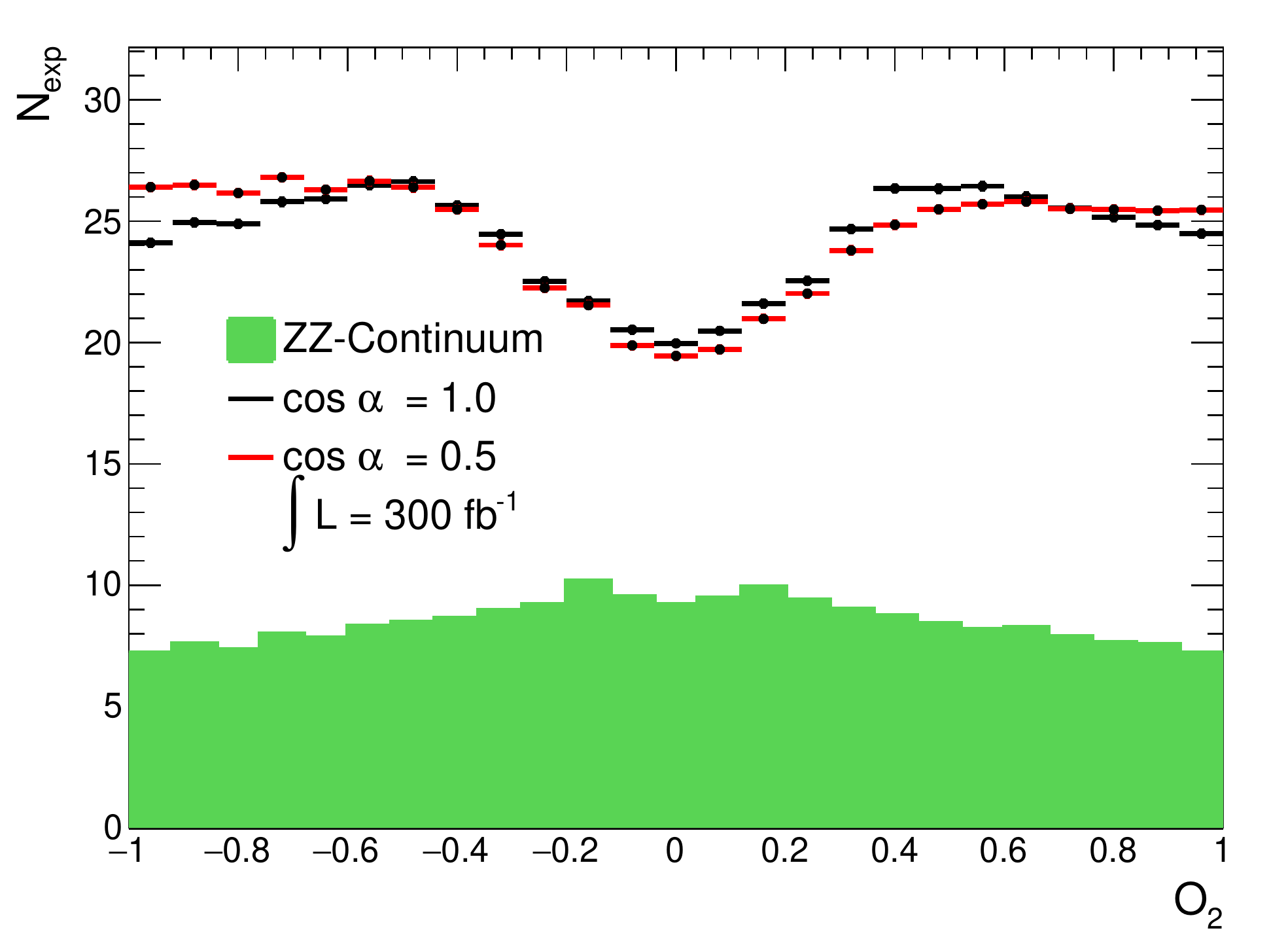}}
\end{minipage}
\hfill
\begin{minipage}[h]{0.49\linewidth}
\center{\includegraphics[width=1\linewidth]{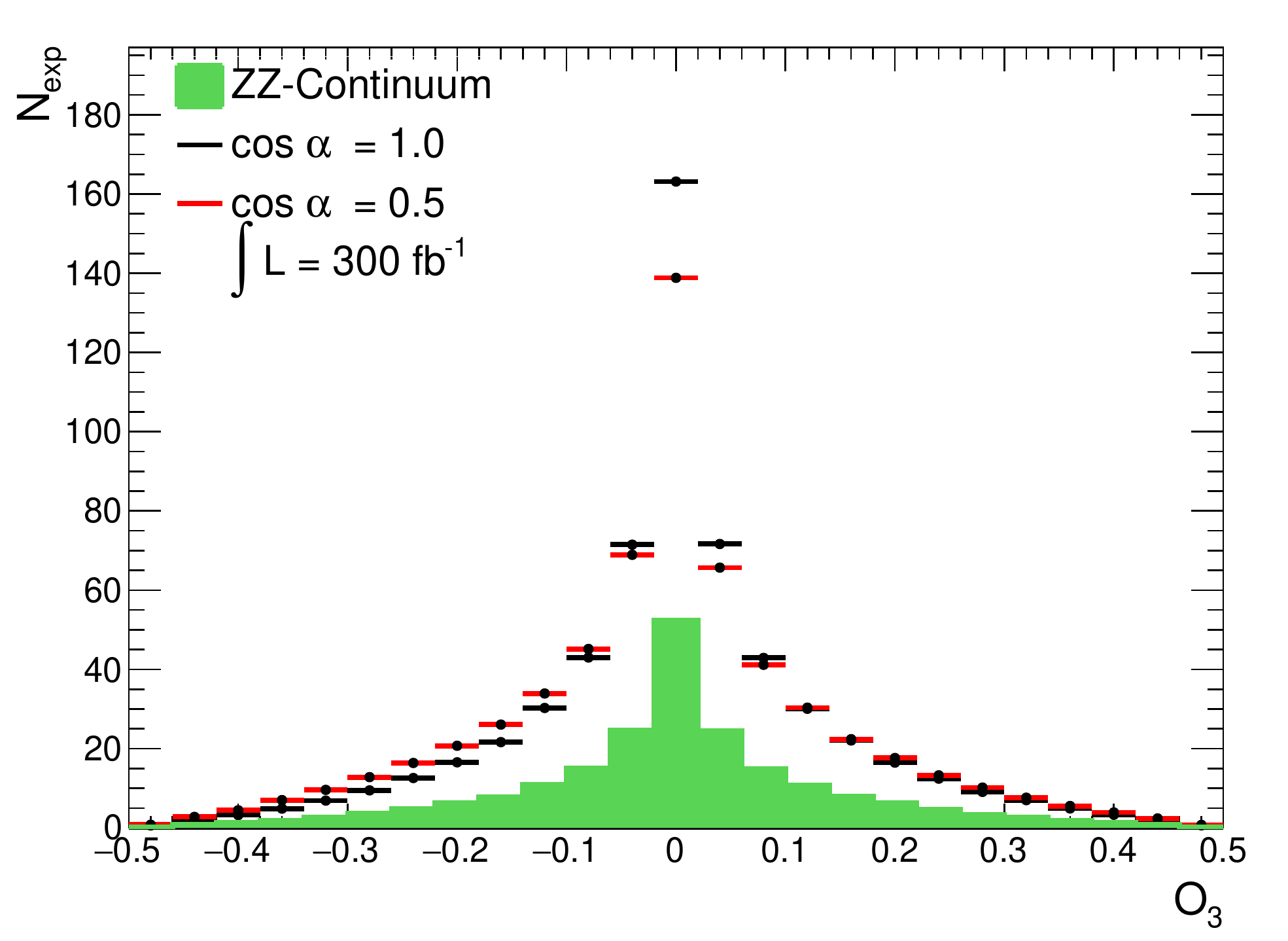}}
\end{minipage}
\vfill
\begin{minipage}[h]{0.49\linewidth}
\center{\includegraphics[width=1\linewidth]{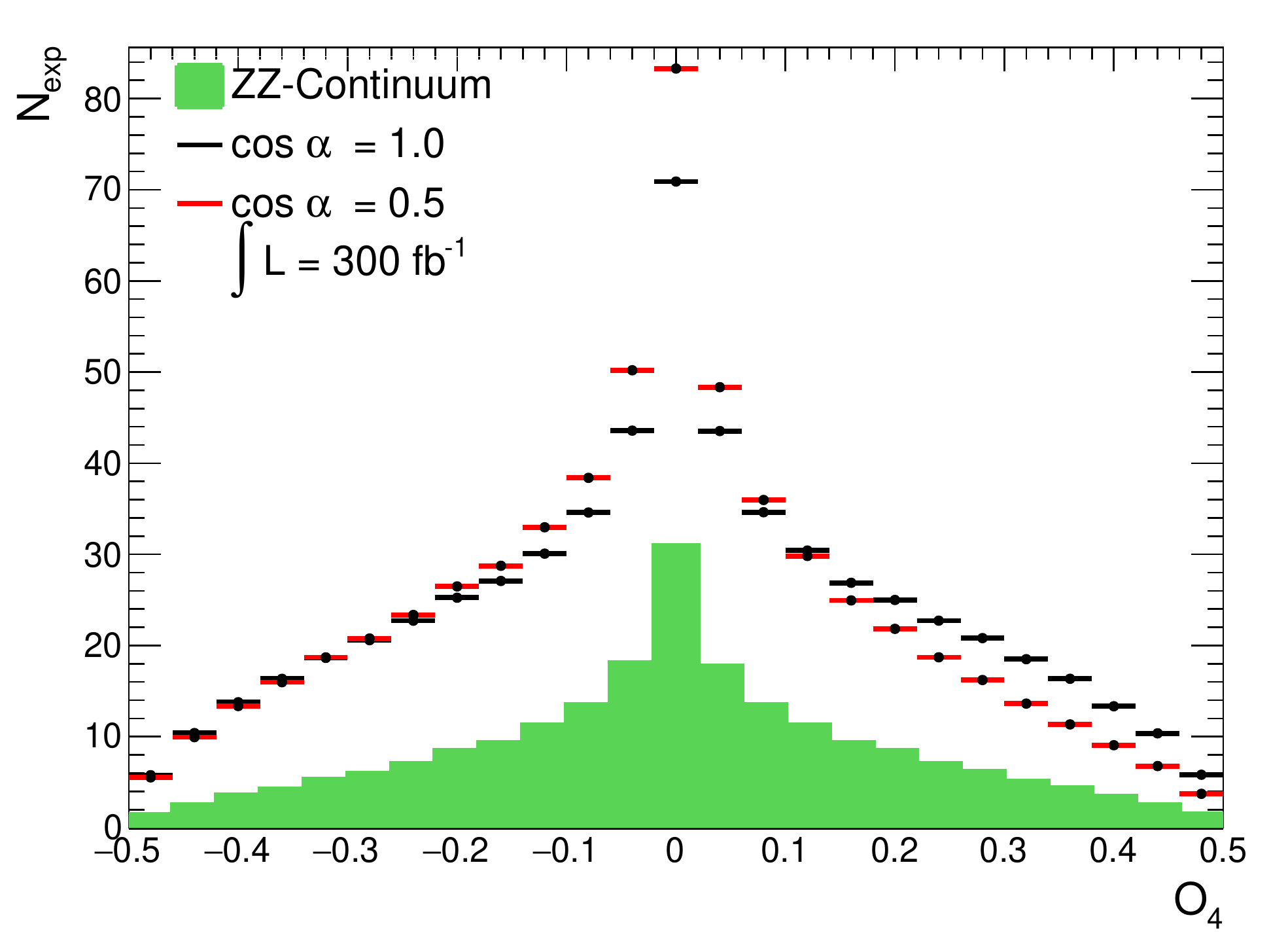}}
\end{minipage}
\hfill
\begin{minipage}[h]{0.49\linewidth}
\center{\includegraphics[width=1\linewidth]{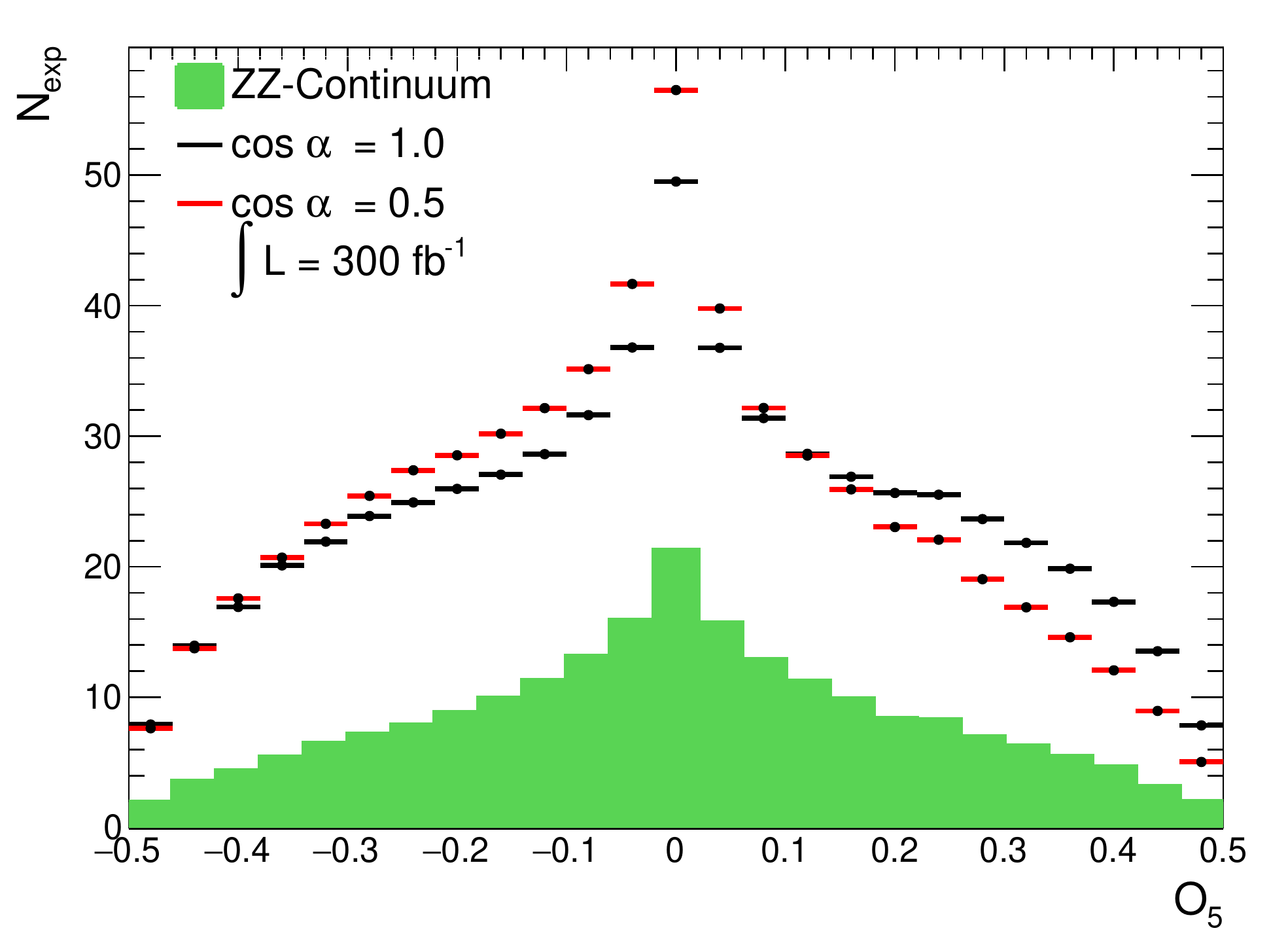}}
\end{minipage}
\caption{Distributions of observables $O_2, O_3, O_4$ and $O_5$ for 
two values of the mixing angle $\alpha$. }
\label{obs_dist}
\end{figure}

\begin{figure}[htbp]
\centering
\includegraphics[width=1.0\columnwidth]{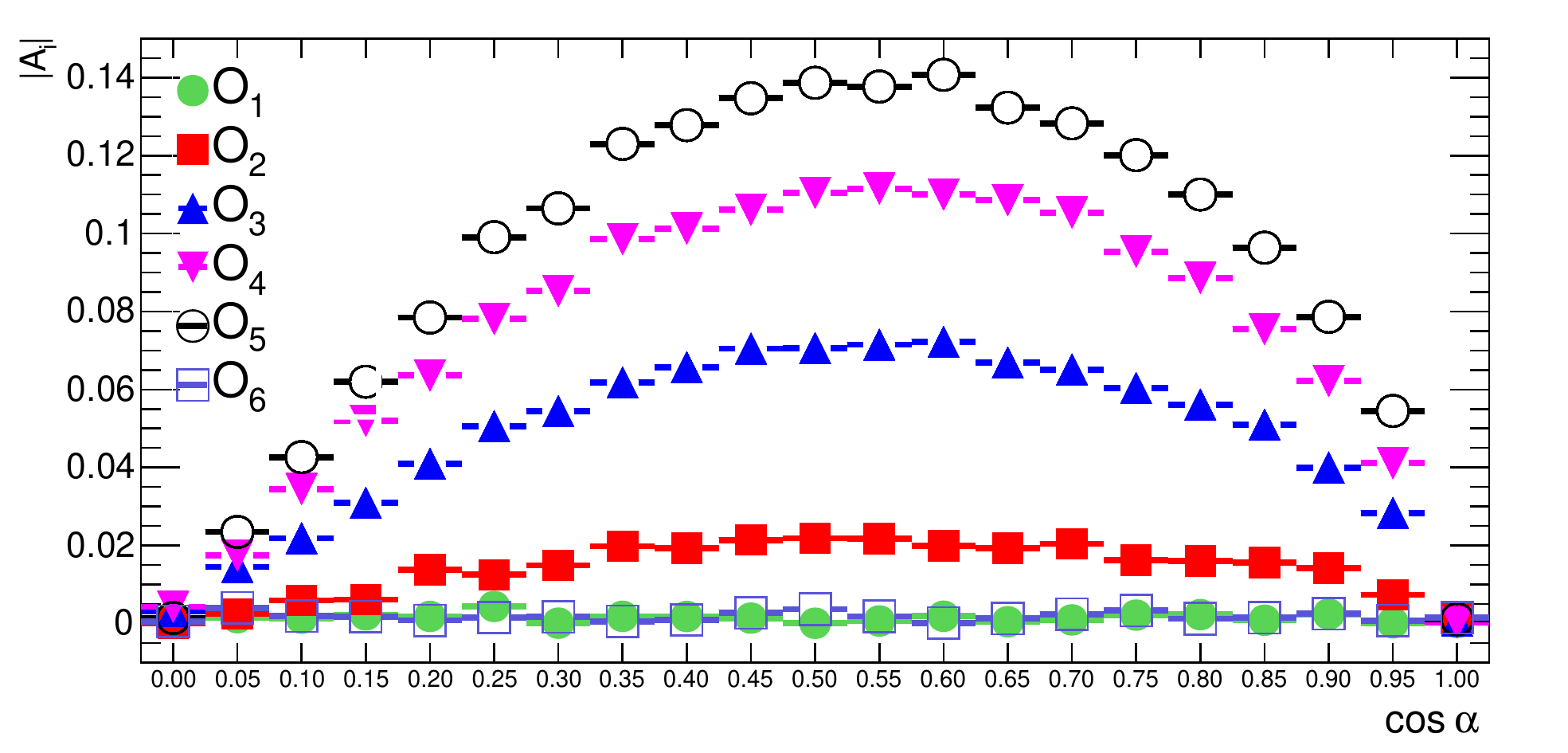}
\caption{Asymmetries generated for observables $O_i$.}           
\label{asym}         
\end{figure}

The distributions of observables  $O_2, O_3, O_4$  and $O_5$ 
for two values of the mixing angle $\cos\alpha=1$ and $\cos\alpha=0.5$
are shown in Fig.~\ref{obs_dist}.
Signal $H\to ZZ\to 4l$  events are generated using the production and decay  model 
defined in Table~\ref{model}. The contributions from the signal and  $q\bar{q}\to ZZ \to  4l$ 
background are normalised to their respective expectations at  $300 fb^{-1}$.     
It is noted that the presence of CP-mixing leads to distortions of distributions of selected 
observables. The distributions of $O_2$ through $O_5$ become asymmetric in the presence
of a real component of $g_4$. This asymmetry is especially pronounced for $O_4$ and $O_5$.
As suggested in \cite{asymmetries}, the background is CP conserving and the 
corresponding distributions of observables are symmetric. The shapes of asymmetries
 $A_i$ for the  model presented in Table~\ref{model} are shown in Fig.~\ref{asym}.
The pure  CP-even and CP-odd cases are given by $\cos\alpha = 1$ and $\cos\alpha = 0$, respectively.

Note, that according to the structure of Lagrangian (Eq.~(\ref{eq:lagrange})) the CP-violating contribution
is defined by the parameter $p=\tilde{K}_{AZZ}\tan\alpha$. This parameter thus determines the corresponding
asymmetries of angular observables. Knowing the distribution of asymmetries for given $\tilde{K}_{AZZ}$
it is possible to obtain the corresponding distributions for any $\tilde{K}_{AZZ}$ by using the condition
$p=const$.

It is noted, that for the physics model used in this study, the observables $O_1$ and $O_6$ do not generate
asymmetries visible with the current Monte Carlo sample. The consistency of these asymmetries with zero confirms that additional effects that are taken into account in our work such as lepton interference, off-shell $ZZ$ production, $ZZ$ background, experimental cuts and detector acceptance do not produce an artificial asymmetry not related with the presence of CP-odd terms. The asymmetric behaviour is clearly visible for
$O_2$ through $O_5$.  The asymmetries for  $O_4$ and $O_5$ calculated using Eq.~(\ref{eq:asym}) 
may exceed $10\%$ .  


In Fig.~\ref{asym} asymmetry plots are given for $\cos\alpha$ in the range from $0$ to $1$. For negative $\cos\alpha$
the asymmetries change sign but keep the same shape. This property allows using the asymmetry approach to measure
the relative phase in the amplitude of Eq.~(\ref{eq:ampl}). 

The significance of the expected asymmetry can be estimated as:
\begin{equation}
S = \Delta N/\sqrt{N} = A_i N_S /\sqrt{N},  \nonumber  
\label{eq:signif}
\end{equation}
where $N = N_S + N_B$ is the total number of  signal  and background events and $\Delta N$ is the difference
in the number of events with  $O_i < 0$ and $O_i > 0$. It is also noted that  $\Delta N \approx \Delta N_S$,
 because the $ZZ$ background does not  contribute to asymmetries at leading order. Following the results of 
 the simulation presented in ~\cite{ATLAS-Collaboration:2013jwa}, the number of signal and background events at 
 $\sqrt{s} = 14$~TeV can be estimated as: $N_S = 1.32 L$ and $N_B = 0.71 L$ respectively. Here $L$ represents the
 integrated luminosity in ${\rm fb}^{-1}$. A dataset with the integrated luminosity of $300\;{\rm fb}^{-1}$ is expected
 to be collected during the Run III of the LHC.

Using the above expressions, one can estimate an expected asymmetry of about  $9.5\%$ to be measured with this data sample.
The corresponding significance will be around two standard deviations.
The region $0.340< \cos\alpha < 0.789$ will then be excluded at $95\%$ CL.

This exclusion range can be expressed in terms of $f_{g_4}$ fraction of events \cite{Khachatryan:2014kca} 
arising from the anomalous coupling $g_4$:
\begin{equation}
f_{g_4} = \frac {\sigma_4 |g_4|^2}{\sigma_1 |g_1|^2 + \sigma_4 |g_4|^2},
\label{eq:fg4}
\end{equation}
where $g_i$ are couplings of the decay vertex, and $\sigma_i$ is the cross section of the processes $H\to ZZ \to 4l $ 
corresponding to $g_i = 1, g_{i \neq j} = 0$. 
 Eq.~(\ref{eq:fg4}) can be rewritten in terms of the mixing angle $\alpha$ as:
\begin{equation}
f_{g_4}=\frac{1}{1+\frac{\sigma_{1}}{\sigma_{4}}\,\left(\frac{k_{SM}}{\tilde K_{AZZ}}\right)^2\cot^{2}\alpha},  \nonumber
\end{equation}
where the ratio of cross sections $\sigma_4/\sigma_1=0.139$ is obtained from the Monte Carlo generator.

The range of the fraction of events of Eq.~(\ref{eq:fg4}) close to $1$ 
has been already excluded by CMS  \cite{Khachatryan:2014kca}. Taking this into account, the 
exclusion limit obtained in the presented analysis becomes
$f_{g_4} < 0.206$ at    $300\; {\rm fb}^{-1}$  
for the model described by the Lagrangian of Eq.~(\ref{eq:lagrange}) and
parameters given in Table~\ref{model}.

For the high luminosity LHC, assuming the same signal and background yields per fb  
as above,  the following exclusion range can be established: 
$0.089 < \cos\alpha < 0.968$ at $95\%$ CL. 
This corresponds to an upper limit $f_{g_4} = 0.028$  at   $3000\;{\rm  fb}^{-1}$.

In the same way as above, we performed estimates for four more values of the model parameter
$\tilde{K}_{AZZ}$. Monte Carlo samples were generated for each point of two dimensional model space 
$(\cos\alpha, \tilde{K}_{AZZ})$. The number of signal events was calculated 
as $N_S=N_S^{SM}\sigma/\sigma^{SM}$ assuming constant K-factors. 
The results are presented in Table~\ref{tab:kvar}.
These limits on $f_{g_4}$  are close to the ones expected in ATLAS \cite{ATLAS-Collaboration:2013jwa} and CMS
\cite{Khachatryan:2014kca} experiments.

\begin{table} [htb!]
\begin{center}
 \begin{tabular}{|c|c|c|c|c|}
\hline
$L,\;{\rm fb}^{-1}$ & \multicolumn{2}{c|}{$300$} &  \multicolumn{2}{c|}{$3000$}\\
	\hline\hline
   $\tilde{K}_{AZZ}/1.76$ & $\Delta c_{\alpha}$ & $f_{g_{4}}$ & $\Delta c_{\alpha}$ & $f_{g_{4}}$   \\
   \hline
   0.6 &      -      &   -   & 0.122-0.921 & 0.026\\
   0.8 & 0.431-0.650 & 0.274 & 0.100-0.953 & 0.027\\
   1.0 & 0.340-0.789 & 0.207 & 0.089-0.968 & 0.028\\
   1.2 & 0.307-0.852 & 0.191 & 0.087-0.975 & 0.031\\
	 1.4 & 0.297-0.886 & 0.188 & 0.086-0.981 & 0.032\\
   \hline
 \end{tabular}
\caption{Upper limit on $f_{g_4}$ and $\cos\alpha$ range excluded at the $95\%$ CL.}\label{tab:kvar}
\end{center}
\end{table}

The region of $\tilde{K}_{AZZ}/1.76$ above 1.4 is not considered. In this region the cross sections exceed the SM 
cross section by more than a factor of two.

\begin{figure}[htbp]
\centering
\includegraphics[width=1.0\columnwidth]{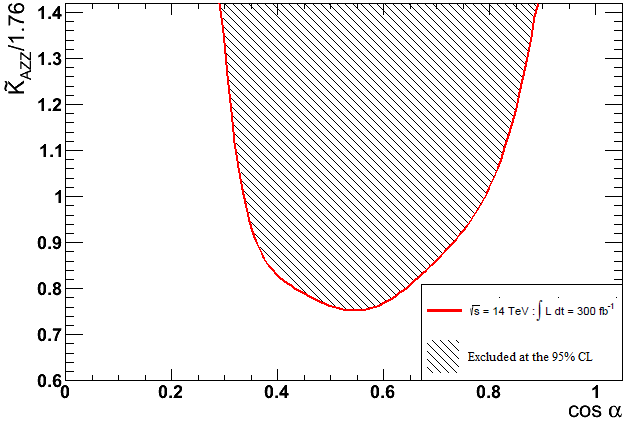}
\caption{The $95\%$ CL exclusion limits for model parameters $c_{\alpha}$, $\tilde{K}_{AZZ}$   
at $300 fb^{-1}$. Regions of rejected model parameters are shown.}
\label{fig:k_vs_cos_300}
\end{figure}

\begin{figure}[htbp]
\centering
\includegraphics[width=1.0\columnwidth]{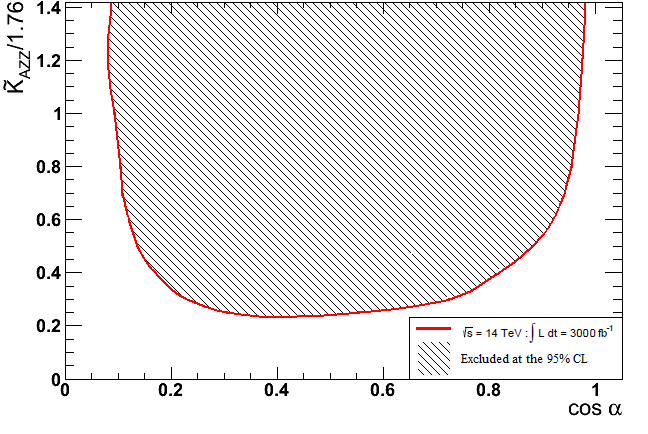}
\caption{The $95\%$ CL exclusion limits for model parameters $c_{\alpha}$, $\tilde{K}_{AZZ}$
at $3000 fb^{-1}$. Regions of rejected model parameters are shown.}
\label{fig:k_vs_cos_3000}
\end{figure}

In Figs.~\ref{fig:k_vs_cos_300} and \ref{fig:k_vs_cos_3000} 
the regions of model parameter space ($c_{\alpha}$, $\tilde{K}_{AZZ}$)  excluded by the current analysis are shown. 
The shadowed areas are excluded at the $95\%$ CL. Lines in Figs.~\ref{fig:k_vs_cos_300} and \ref{fig:k_vs_cos_3000}
represent a polinomial fit to the results of the method of asymmetries. 

Note that CP-odd observables were studied also in \cite{Buchalla:2013mpa}. According to this  article the detection of CP-violating effects is out of reach of the LHC. However, as was mentioned in \cite{Buchalla:2013mpa}, these effects might in principle attain large  values because of numerical enhancements.

\label{sec:fit}
\section{Mixing angle observable fit}
The asymmetries $A_i$ discussed in the Section~\ref{sec:asym}  are integrated 
quantities of  angular observables $O_i$ and thus provide limited information about 
the anomalous contributions to the $HZZ$ vertex.
The optimal sensitivity to these contributions can be obtained by studying the shapes of  distributions of observables 
$O_i$ and their correlations.

\begin{figure}[ttbp]
\center{\includegraphics[width=1.0\linewidth]{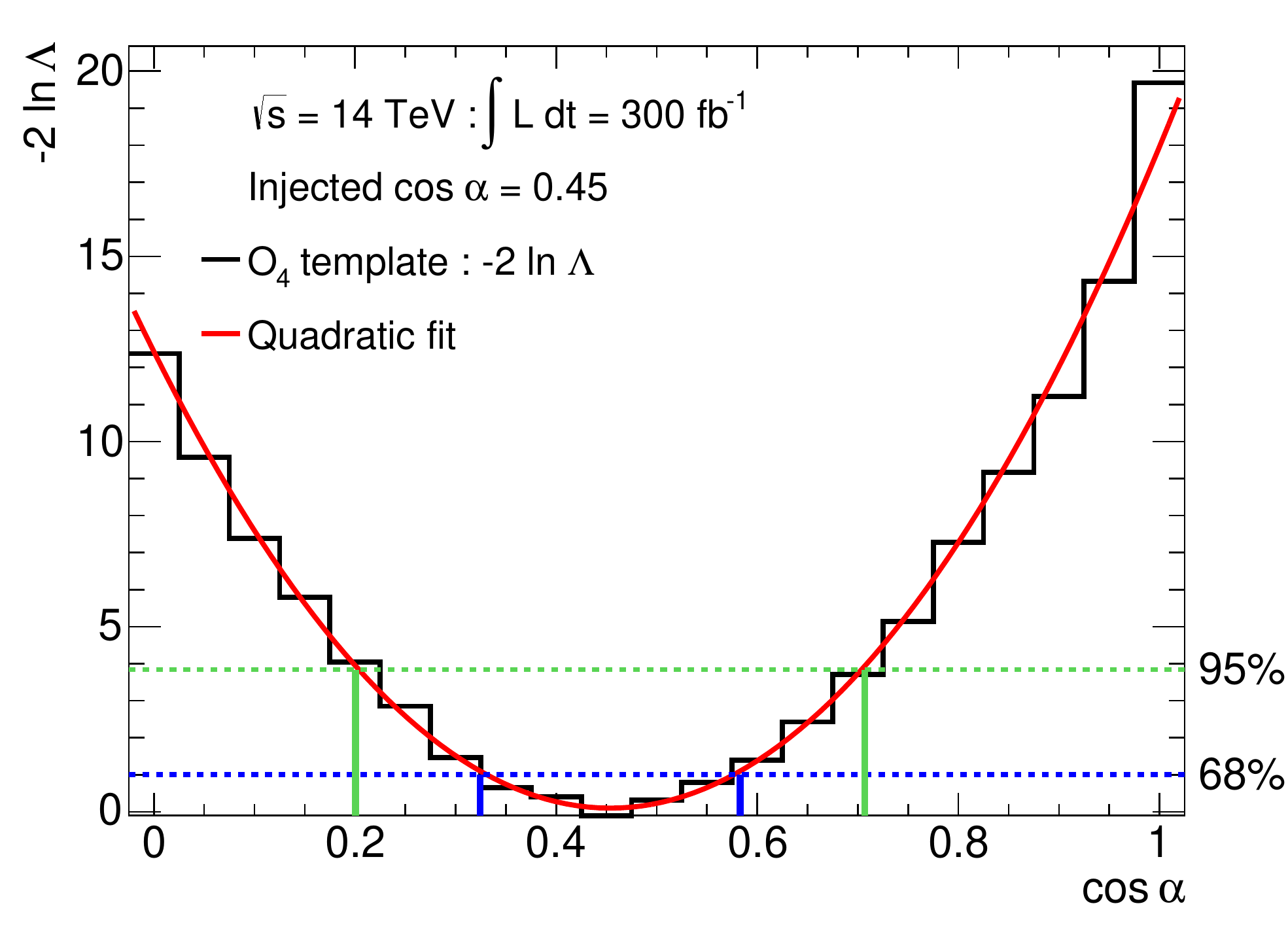}}
\caption{Example of the likelihood curve for the mixing angle observable fit of $O_4$. The definitions of the  $64\%$~CL and 
$95\%$~CL exclusion regions are demonstrated.}
\label{fig:nll}
\end{figure}

The sensitivity of individual observables to the presence of anomalous contributions to the $HZZ$ vertex
is studied by fitting the shape of these observables as a function of the mixing angle. 
The likelihood function of the fit is defined as:
\begin{equation*}
\mathcal{L}(\cos{\alpha}, \mu, \theta) = \prod_{j}^{\mathrm{N_{chan}}} \prod_{i}^{\mathrm{N_{bin}}} P(N_{i,j} | \mu_{j} \cdot S_{i,j}(\cos{\alpha}, \theta) + B_{i,j}(\theta)). 
\end{equation*}
Here besides the parameter of interest $\cos{\alpha}$, two nuisance parameters have been introduced: the best fitting signal strength $\mu$ 
and a systematic normalization uncertainty $\theta$.  The likelihood function is a product over the different final states and bins of the 
specific observable that is being fitted. In each bin, the observed number of events from pseudo-data $N$, is compared to the expected number
of events of the model $S+B$ assuming a Poissonian distribution of entries $P$. By varying the mixing parameter $\cos{\alpha}$ of the likelihood for a 
given dataset we can construct the standard log-likelihood test statistic:
\begin{equation*}
  -2 \ln \Lambda( \cos{\alpha} ) = -2 \ln \frac{\mathcal{L}(\cos{\alpha})}{\mathcal{L}(\cos{\hat{\alpha}})},
\end{equation*} 
where $\hat{\alpha}$ denotes the mixing angle that maximises the likelihood function over the scan. The other likelihood parameters are profiled 
at the corresponding $\cos{\alpha}$ value. The 95$\%$ exclusion is reached when $-2 \ln \Lambda( \cos{\alpha} ) > 3.84$.
The definitions of the  $64\%$~CL and $95\%$~CL exclusion regions is demonstrated in Fig.~\ref{fig:nll}.

\begin{figure*}[t]
\begin{minipage}[h]{0.31\linewidth}
\center{\includegraphics[width=1\linewidth]{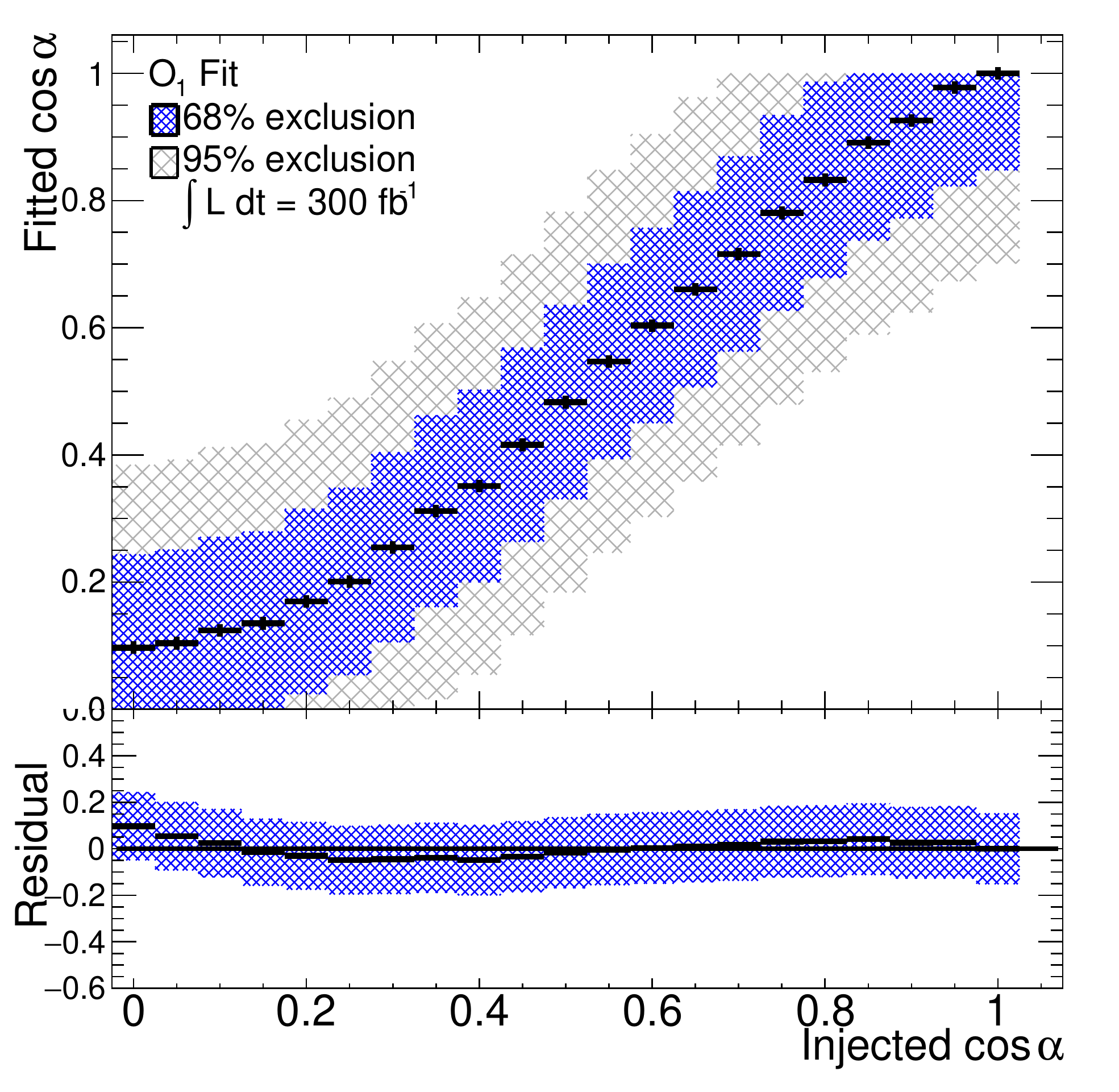}}  
\end{minipage}
\hfill
\begin{minipage}[h]{0.31\linewidth}
\center{\includegraphics[width=1\linewidth]{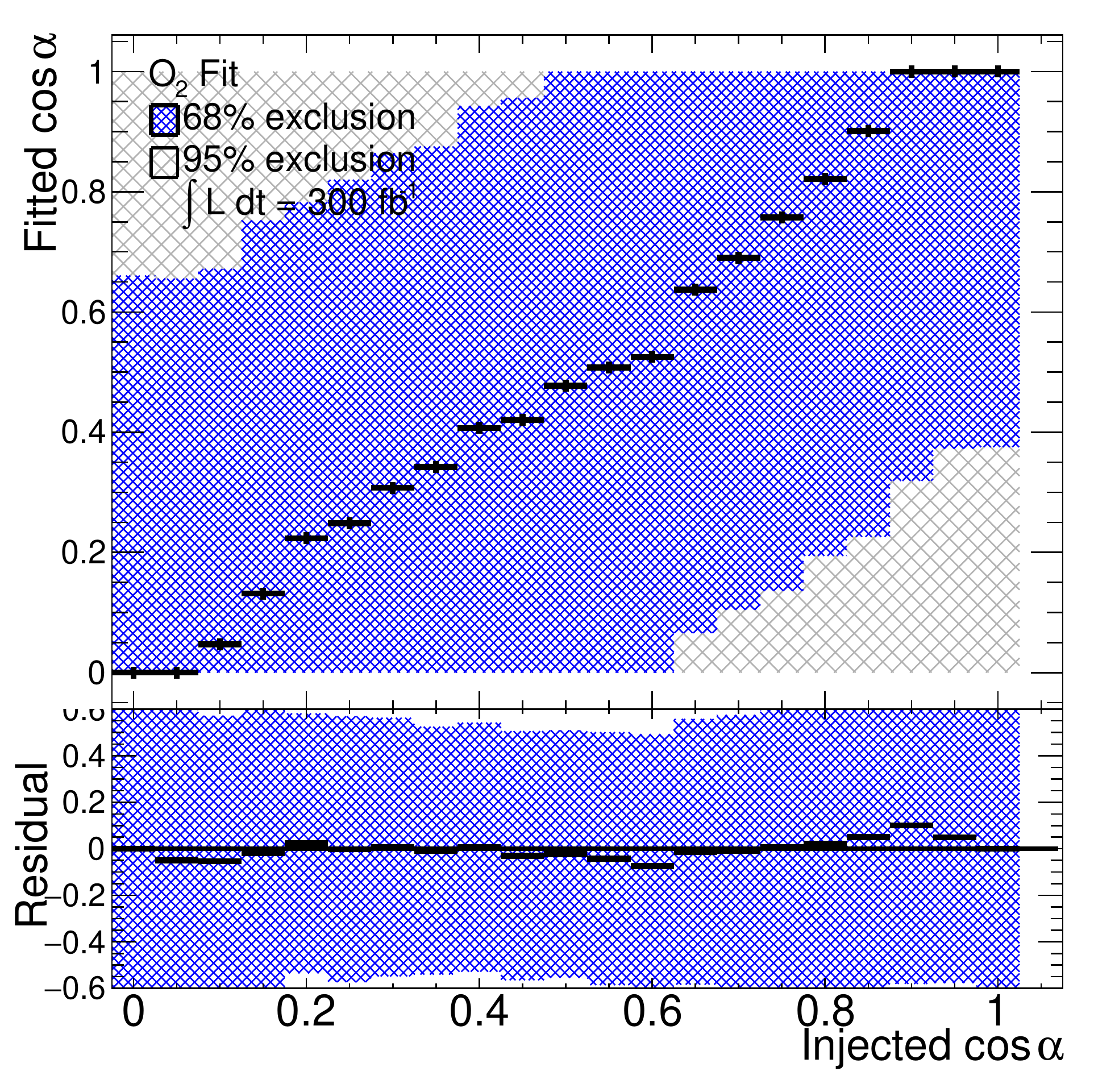}}  
\end{minipage}
\hfill
\begin{minipage}[htbp]{0.31\linewidth}
\center{\includegraphics[width=1\linewidth]{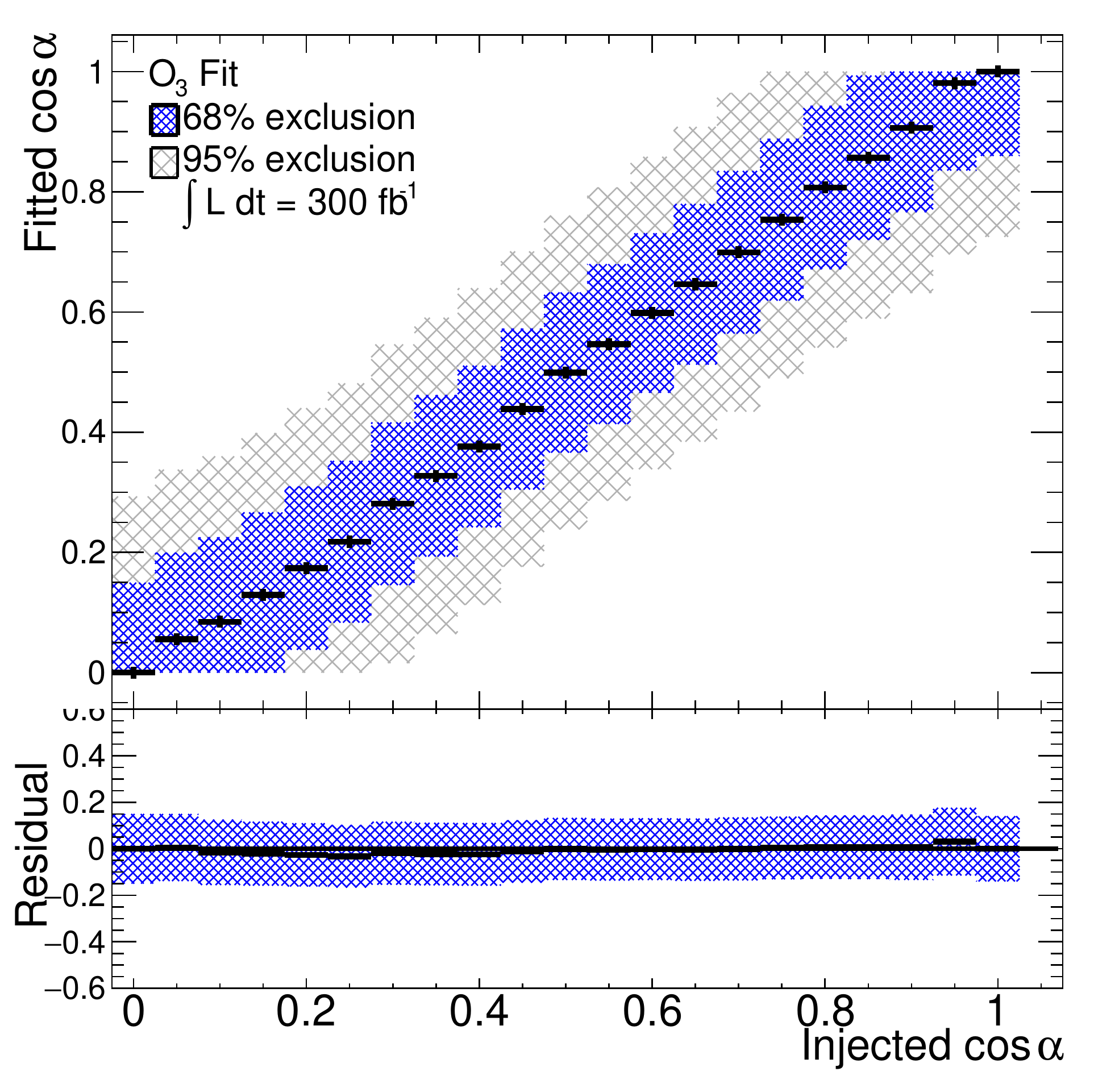}}  
\end{minipage}
\vfill
\begin{minipage}[h]{0.31\linewidth}
\center{\includegraphics[width=1\linewidth]{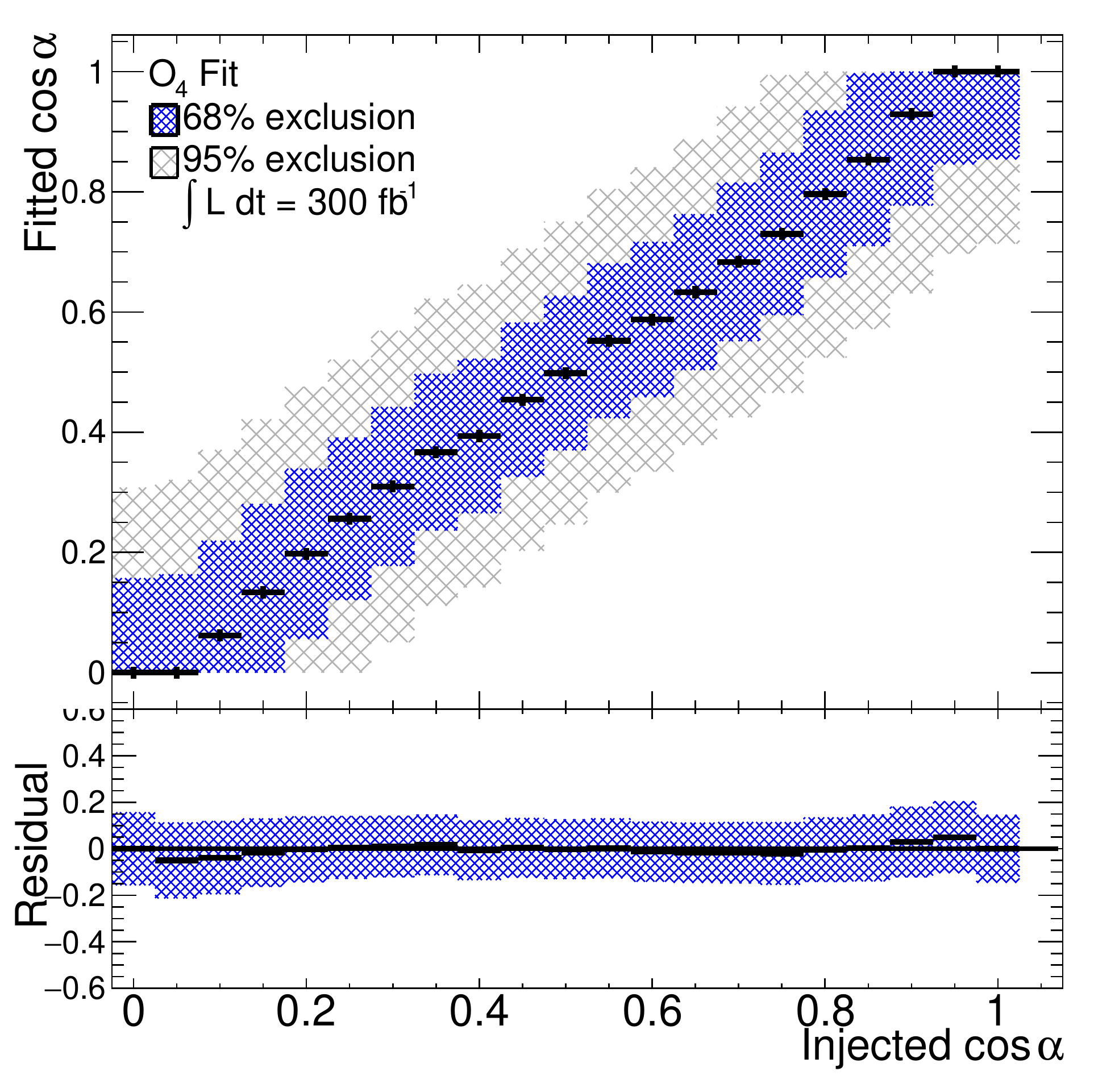}}  
\end{minipage}
\hfill
\begin{minipage}[h]{0.31\linewidth}
\center{\includegraphics[width=1\linewidth]{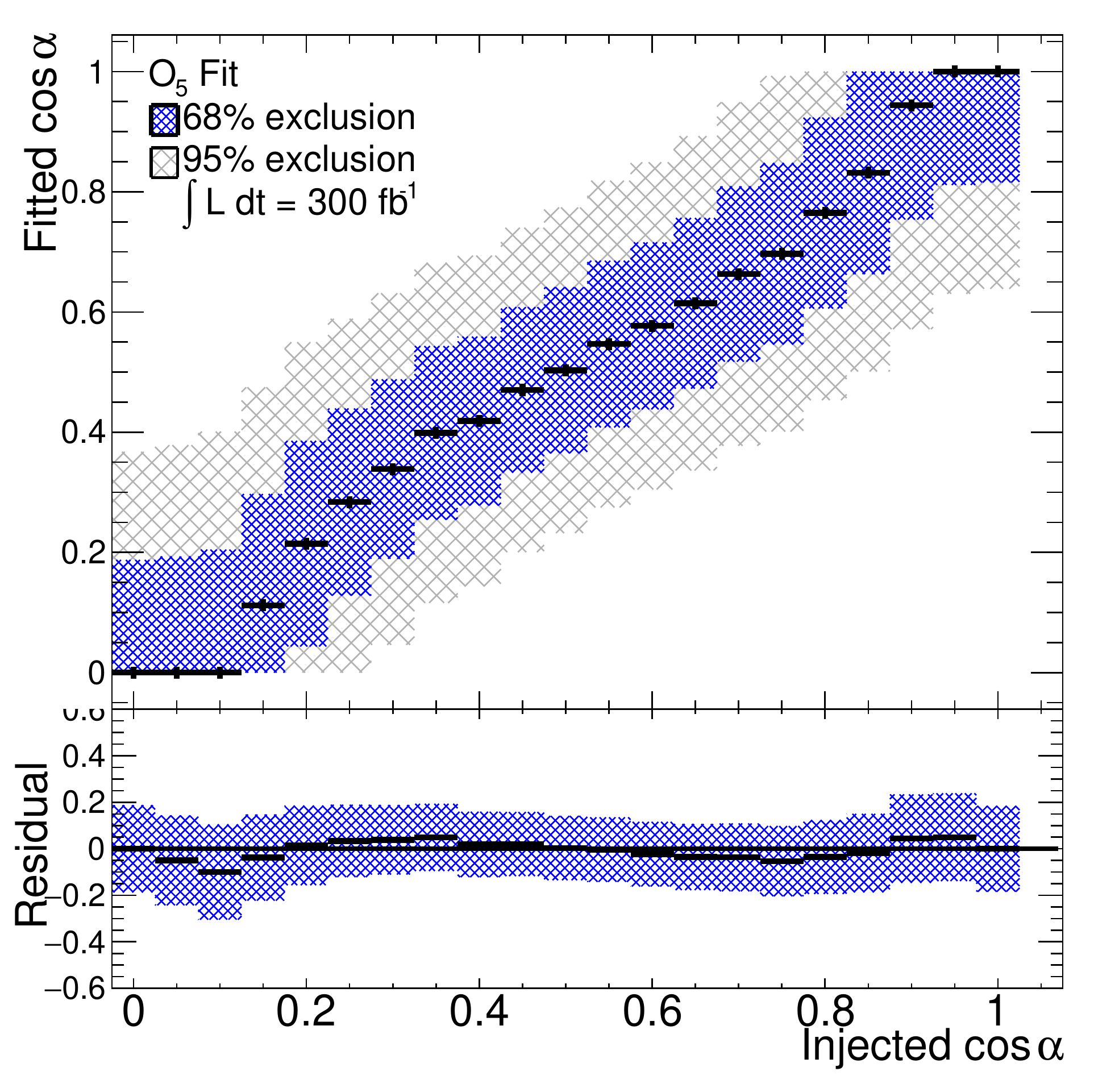}}  
\end{minipage}
\hfill
\begin{minipage}[h]{0.31\linewidth}
\center{\includegraphics[width=1\linewidth]{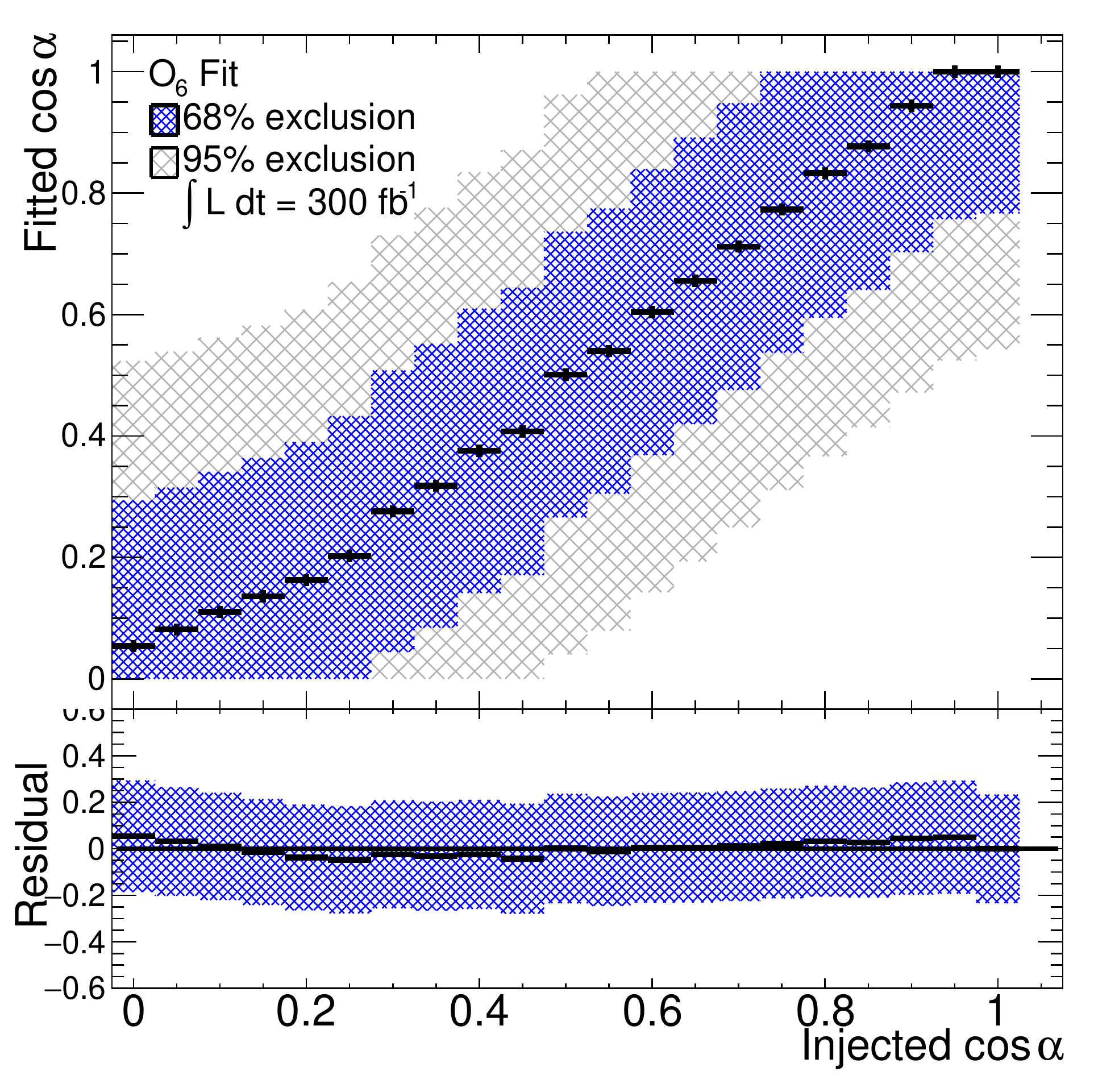}}  
\end{minipage}
\caption{Results of  the mixing angle $\alpha$ scan produced with the mixing angle observable fit 
corresponding to the integrated luminosity of $300\;{\rm fb}^{-1}$.}
\label{fig:observables_1}
\end{figure*}

Results of the scan of  the mixing angle $\alpha$ produced with the mixing angle observable fit 
corresponding to the integrated luminosity of $300\;{\rm fb}^{-1}$ are presented 
in Fig.~\ref{fig:observables_1}.
The results are reported for the model with $\tilde{K}_{AZZ} =1.76$ and remaining 
parameters as defined in Table~\ref{model}. The values of the mixing angle $\cos\alpha$ used 
to generate the input pseudo-data are marked on the $x$-axis. Every bin of the injected $\cos \alpha$ 
on represents the null-hypothesis likelihood curve similar to Fig.~\ref{fig:nll}.  The $y$-axis shows the 
$\cos\hat{\alpha}$ values reconstructed in the fit. The blue and grey dashed areas represent the $64\%$~CL 
and $95\%$~CL limits respectively. The white area in each bin of injected $\cos\alpha$ is excluded at 
$95\%$~CL. As expected,  the sensitivity to the mixing angle varies for different observables,
resulting in significantly different exclusion regions. The weakest exclusion is reached with the 
$O_{2}$, while the strongest is reached with the $O_{4}$.

\begin{figure*}[htbp]
\begin{minipage}[h]{0.31\linewidth}
\center{\includegraphics[width=1\linewidth]{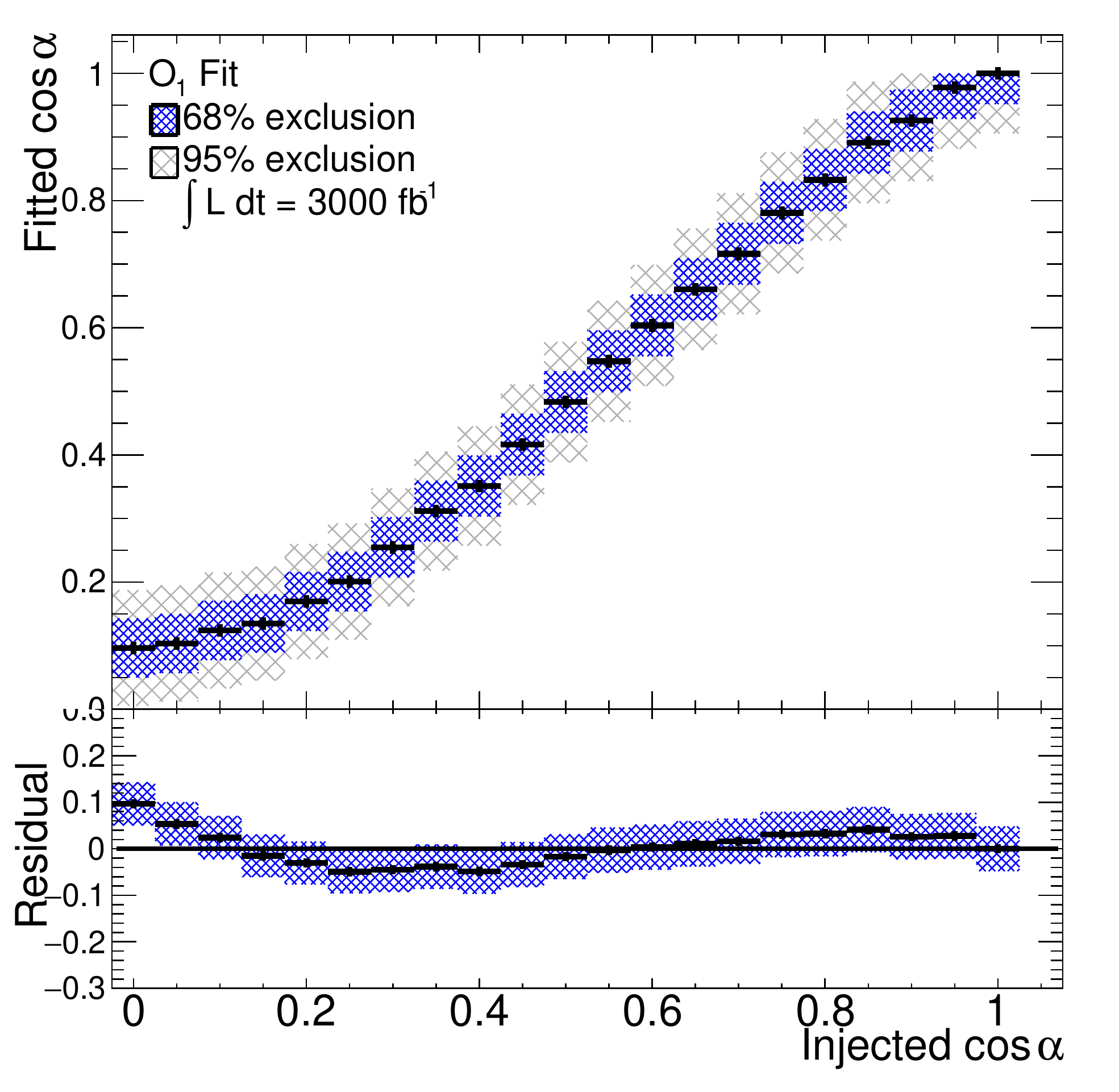}} 
\end{minipage}
\hfill
\begin{minipage}[h]{0.31\linewidth}
\center{\includegraphics[width=1\linewidth]{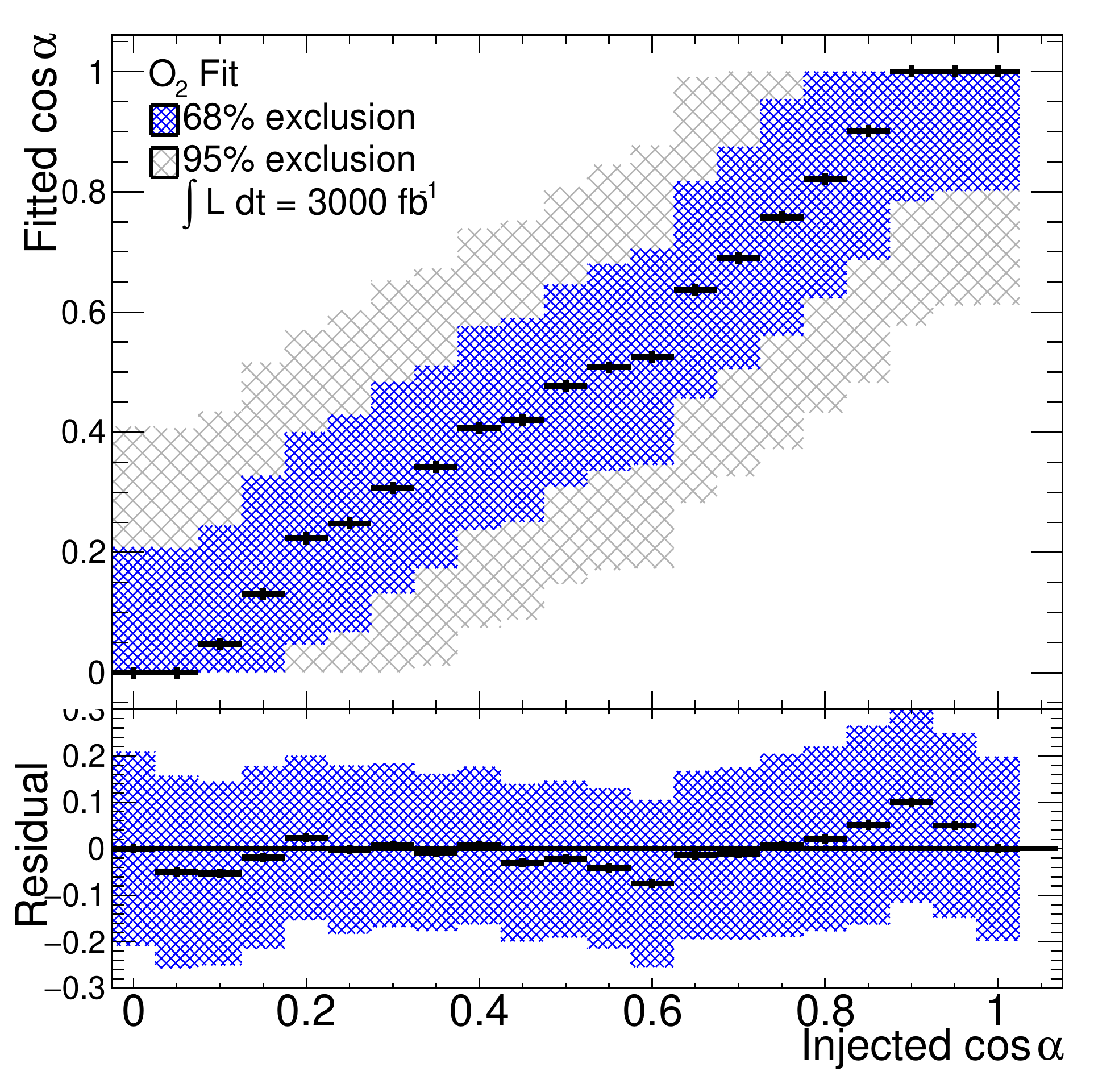}} 
\end{minipage}
\hfill
\begin{minipage}[h]{0.31\linewidth}
\center{\includegraphics[width=1\linewidth]{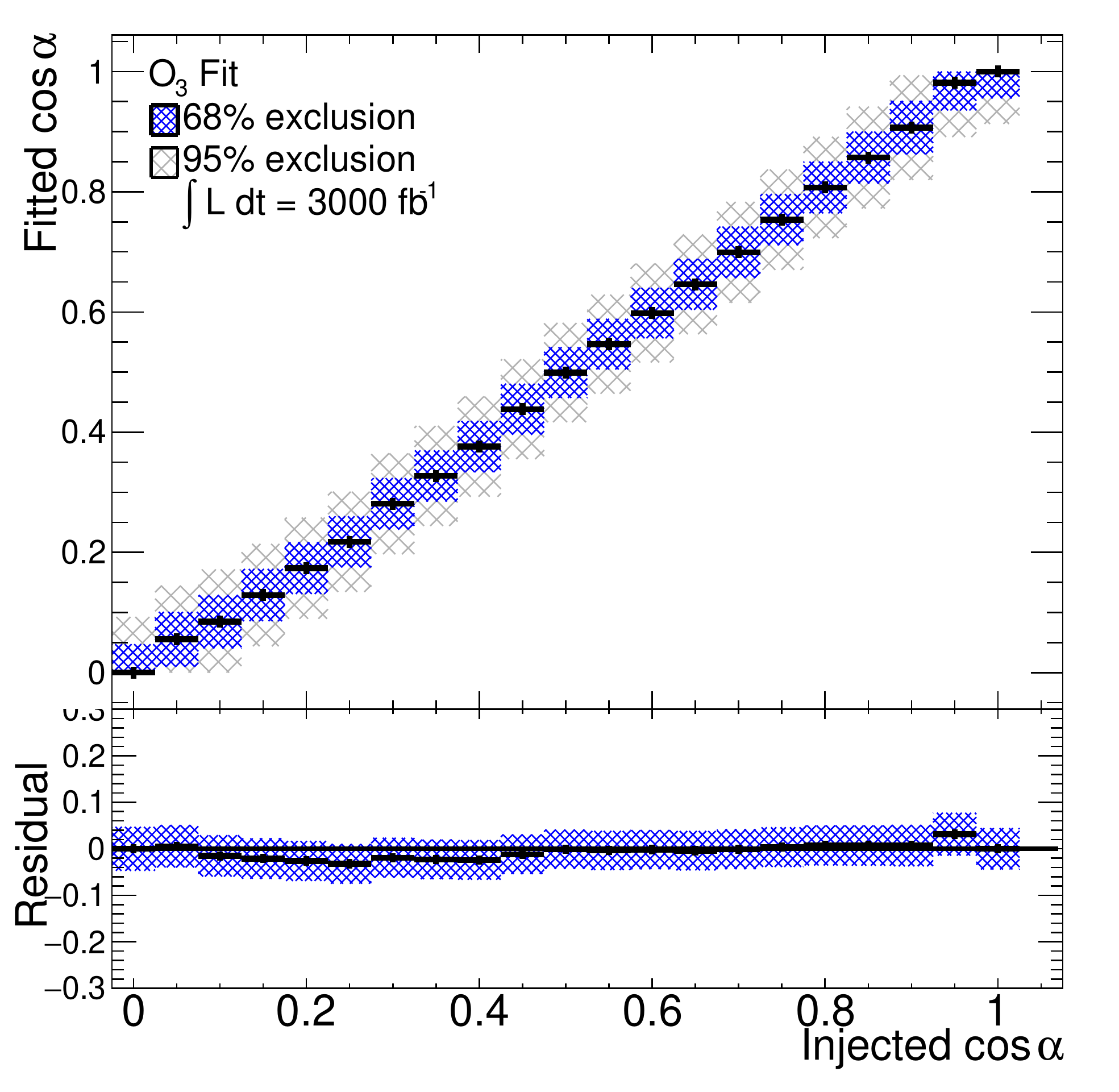}} 
\end{minipage}
\vfill
\begin{minipage}[h]{0.31\linewidth}
\center{\includegraphics[width=1\linewidth]{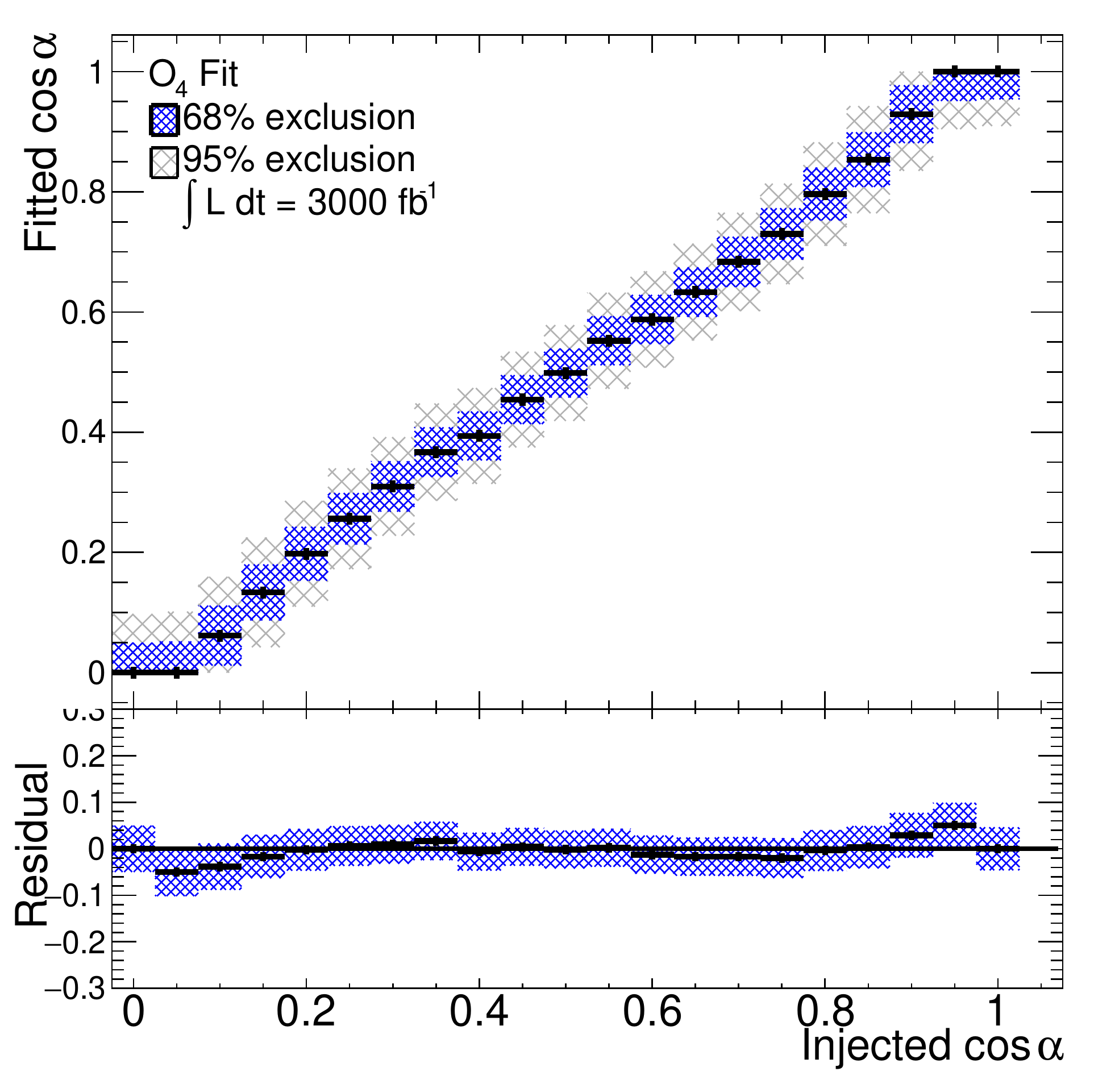}} 
\end{minipage}
\hfill
\begin{minipage}[h]{0.31\linewidth}
\center{\includegraphics[width=1\linewidth]{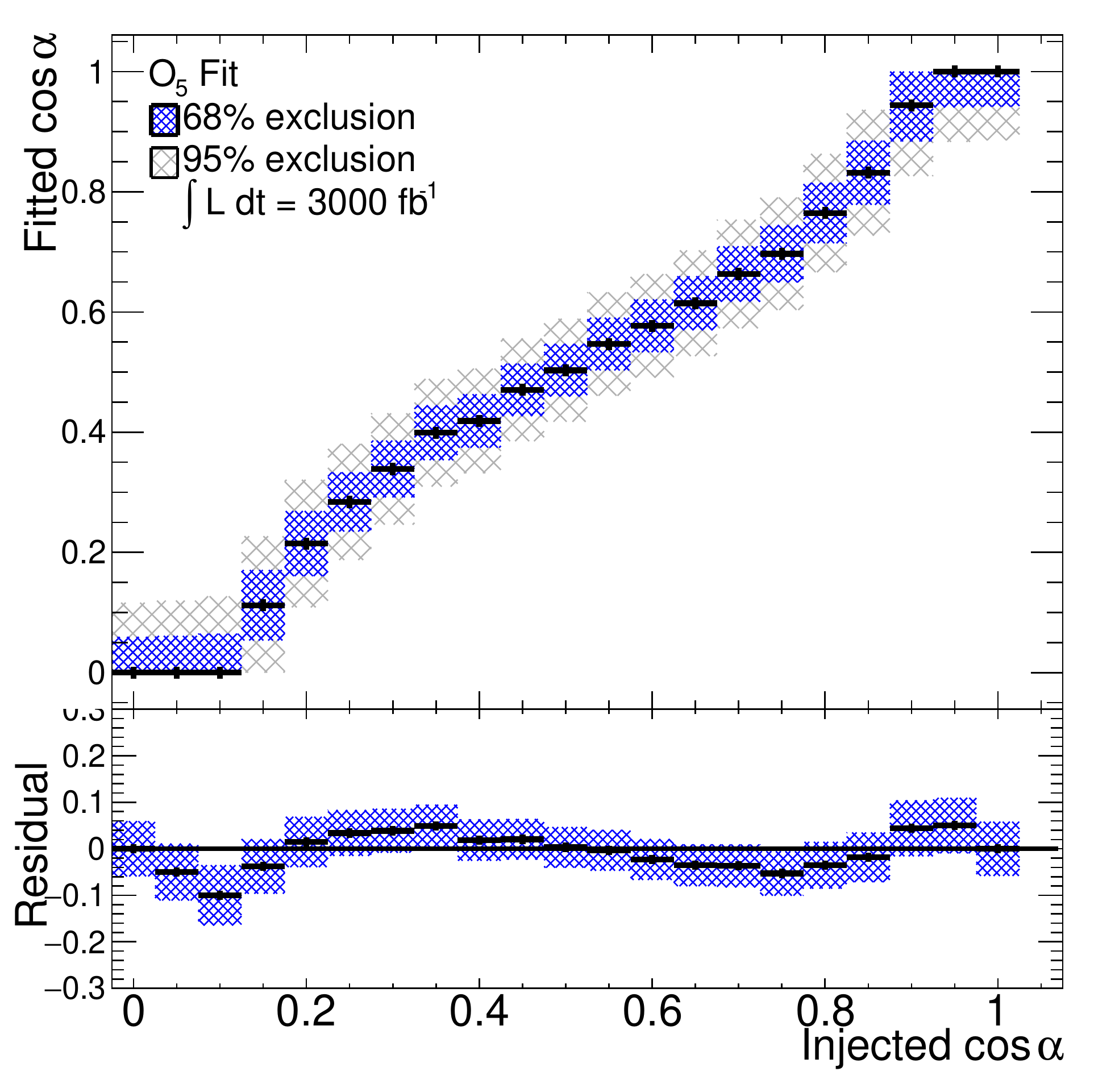}} 
\end{minipage}
\hfill
\begin{minipage}[h]{0.31\linewidth}
\center{\includegraphics[width=1\linewidth]{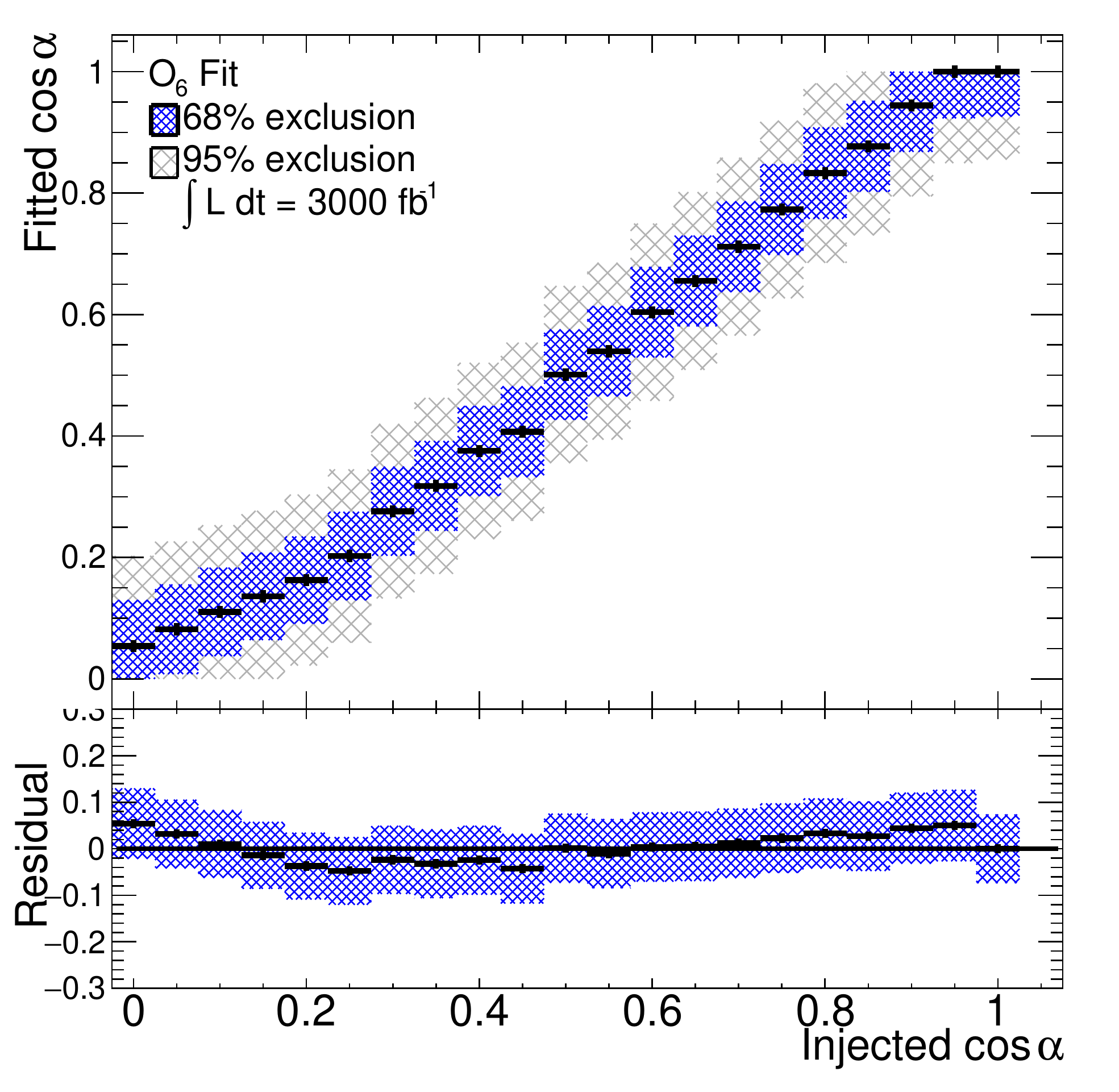}} 
\end{minipage}
\caption{Results of  the mixing angle $\alpha$ scan produced with the mixing angle observable fit 
corresponding to the integrated luminosity of $3000\;{\rm fb}^{-1}$.}
\label{fig:observables_2}
\end{figure*}

The results corresponding to the integrated luminosity of $3000\;{\rm fb}^{-1}$ are presented 
in Fig.~\ref{fig:observables_2}.
Compared to $300\;{\rm fb}^{-1}$, the  $95\%$~CL exclusion regions around the fitted $\cos\hat{\alpha}$
values are significantly reduced. Assuming the pure  Standard Model signal, the following 
exclusion limits can be set using the $O_4$ observable alone: $0<\cos \hat{\alpha}<0.708$ at the $95\%$~CL for 
 $300\;{\rm fb}^{-1}$ and $0<\cos \hat{\alpha}< 0.908$ at the $95\%$~CL for $3000\;{\rm fb}^{-1}$. The exclusion limits 
 obtained from other observables  assuming the Standard Model signal are reported in Table~\ref{tab:ObsFitRes}.

 The exclusion limits obtained for hypothetical BSM signals can be read from Fig.~\ref{fig:observables_1}
 and \ref{fig:observables_2}. It is noted that by fitting the shape of the $O_4$ observable alone the exclusion 
 limits similar to those reported in the Section~\ref{sec:asym} can be obtained. Further improvements can be 
 obtained by combining several observables in the same fit. 
 \begin{table}[h]
   \begin{center}
     \begin{tabular}{|c|c|c|c|c|}
       \hline
       $L,\;{\rm fb}^{-1}$ & \multicolumn{2}{c|}{$300$} &  \multicolumn{2}{c|}{$3000$}   \\
       \hline\hline
       Observable & $\Delta c_{\alpha}$ & $f_{g_{4}}$ & $\Delta c_{\alpha}$ & $f_{g_{4}}$   \\
       \hline
       $O_1$  & 0 - 0.695  & 0.315 & 0 - 0.903  &  0.089 \\
       $O_2$  & -          & -     & 0 - 0.604  &  0.428 \\
       $O_3$  & 0 - 0.719  & 0.287 & 0 - 0.911  &  0.081 \\
       $O_4$  & 0 - 0.708  & 0.300 & 0 - 0.908  &  0.084 \\
       $O_5$  & 0 - 0.631  & 0.394 & 0 - 0.883  &  0.108 \\
       $O_6$  & 0 - 0.533  & 0.520 & 0 - 0.852  &  0.104 \\
       \hline
     \end{tabular}
     \caption{\label{tab:ObsFitRes}Upper limit on $f_{g_4}$ and $\cos\alpha$ range excluded at the $95\%$ CL
       with the mixing angle observable fit. The Standard Model signal is assumed. The BSM templates
       are generated according to the model defined in Table~\ref{model} with $\tilde{K}_{AZZ} =1.76$.}
   \end{center}
 \end{table}
%
%

\label{sec:conc}
\section{Conclusion}
In this article  studies of  tensor structure of the $HZZ$ vertex are presented.
The investigation is performed using the $pp \to H \to ZZ \to 4l$ process assuming the 
gluon fusion production of the spin-0 resonance.  The background contributions, detector resolution, 
trigger and selection efficiencies expected  for the LHC are taken into account.
Two different approaches  to detect CP-violation effects in the $HZZ$ vertex were used. The first approach is based on 
a simple counting experiment for angular asymmetries of CP-sensitive observables. It was shown that the presence of 
CP  violating terms may result in angular asymmetries exceeding $10 \%$. 
The $95 \%$~CL exclusion ranges for the mixing angle at different parameters of spin-0 
Higgs boson model including the Standard Model CP-even term and anomalous CP-odd term $g_4$
are calculated. These results are also presented in terms of the effective cross section fraction $f_{g_4}$. 
The obtained limits  are comparable with the ATLAS and CMS projections for Run III at the LHC 
and the High-Luminosity LHC presented in ~\cite{Khachatryan:2014kca, ATLAS-Collaboration:2013jwa}.   
 
The sensitivity of individual observables to the presence of anomalous contributions to the 
$HZZ$ vertex was studied by fitting the shape of these observables as a
function of the mixing angle. It is demonstrated that using a single most sensitive 
observable, this approach gives $f_{g_4}$ limits comparable with asymmetries method and
 with the ATLAS and CMS projections. Compared to the method of angular asymmetries, 
 this approach has an advantage of using the complete  shape information of  CP-odd observables.
It is demonstrated that some of the observables, that do not generate  significant angular asymmetry
in presence of significant CP-mixing, can still provide restrictive $f_{g_4}$ limits when their complete shape
is analysed. Combining several CP-odd observables in the same fit or combining several angular 
asymmetries would likely further improve sensitivity  to the CP violating coupling. 
It is noted that careful experimental investigation of all observables, even not the leading ones, 
is important, since they probe different terms of the $HZZ$ vertex. 
\section*{ACKNOWLEDGMENTS}
We would like to  thank our ATLAS and CMS colleagues for many fruitful discussions and suggestions.
We also thank MadGraph and JHU teams for many useful advices. We are grateful to
A.~Mincer for reading the manuscript and providing valuable suggestions.
We thank A.~Baumgartner, D.~Gray, and T.~Reid for their help with MC samples production.
The work of R.~Konoplich is partially supported  by the US National 
Science Foundation under Grants No.PHY-1205376 and No.PHY-1402964.
The work of K.~Prokofiev is partially supported by a grant from the Research 
Grant Council of the Hong Kong Special Administrative Region, China (Project 
Nos. CUHK4/CRF/13G).

\bibliography{main}

\begin{thebibliography}{10}

\bibitem{cms_higgs}
{CMS Collaborations},
\newblock {Phys. Lett. B} {\bf {716}}, {30} ({2012}).

\bibitem{atlas_higgs}
{ATLAS Collaborations},
\newblock {Phys. Lett. B} {\bf {716}}, {1} ({2012}).

\bibitem{atlas_couplings}
{ATLAS Collaborations},
\newblock {Phys. Lett. B} {\bf {726}}, {88} ({2013}).

\bibitem{Khachatryan:2014kca}
{CMS Collaboration},
\newblock (2014), arXiv:1411.3441.

\bibitem{Chatrchyan:2012jja}
{CMS Collaboration},
\newblock Phys. Rev. Lett. {\bf 110}, 081803 (2013).

\bibitem{atlas_spin}
{ATLAS Collaborations},
\newblock {Phys. Lett. B} {\bf {726}}, {120} ({2013}).

\bibitem{Chang:1993jy}
D.~Chang, W.-Y. Keung, and I.~Phillips,
\newblock Phys. Rev. D {\bf 48}, 3225 (1993).

\bibitem{Grzadkowski:1995rx}
B.~Grzadkowski and J.~Gunion,
\newblock Phys. Lett. B {\bf 350}, 218 (1995).

\bibitem{Gunion:1996vv}
J.~F. Gunion, B.~Grzadkowski, and X.-G. He,
\newblock Phys. Rev. Lett. {\bf 77}, 5172 (1996).

\bibitem{Grzadkowski:1999ye}
B.~Grzadkowski, J.~F. Gunion, and J.~Kalinowski,
\newblock Phys. Rev. D {\bf 60}, 075011 (1999).

\bibitem{Grzadkowski:2000hm}
B.~Grzadkowski, J.~F. Gunion, and J.~Pliszka,
\newblock Nucl. Phys. {\bf B583}, 49 (2000).

\bibitem{Plehn:2001nj}
T.~Plehn, D.~L. Rainwater, and D.~Zeppenfeld,
\newblock Phys. Rev. Lett. {\bf 88}, 051801 (2002).

\bibitem{Choi:2002jk}
S.~Choi, D.~Miller, M.~Muhlleitner, and P.~Zerwas,
\newblock Phys. Lett. B {\bf 553}, 61 (2003).

\bibitem{Buszello:2002uu}
C.~Buszello, I.~Fleck, P.~Marquard, and J.~van~der Bij,
\newblock {Eur. Phys. J. C} {\bf 32}, 209 (2004).

\bibitem{Hankele:2006ma}
V.~Hankele, G.~Klamke, D.~Zeppenfeld, and T.~Figy,
\newblock Phys. Rev. D {\bf 74}, 095001 (2006).

\bibitem{Godbole:2007cn}
R.~M. Godbole, D.~Miller, and M.~M. Muhlleitner,
\newblock JHEP {\bf 0712}, 031 (2007).

\bibitem{Keung:2008ve}
W.-Y. Keung, I.~Low, and J.~Shu,
\newblock Phys. Rev. Lett. {\bf 101}, 091802 (2008).

\bibitem{Berge:2008dr}
S.~Berge and W.~Bernreuther,
\newblock Phys. Lett. B {\bf 671}, 470 (2009).

\bibitem{Cao:2009ah}
Q.-H. Cao, C.~Jackson, W.-Y. Keung, I.~Low, and J.~Shu,
\newblock Phys. Rev. D {\bf 81}, 015010 (2010).

\bibitem{DeRujula:2010ys}
A.~De~Rujula, J.~Lykken, M.~Pierini, C.~Rogan, and M.~Spiropulu,
\newblock Phys. Rev. D {\bf 82}, 013003 (2010).

\bibitem{Gao:2010qx}
Y.~Gao {\em et~al.},
\newblock Phys. Rev. D {\bf 81}, 075022 (2010).

\bibitem{Berge:2011ij}
S.~Berge, W.~Bernreuther, B.~Niepelt, and H.~Spiesberger,
\newblock Phys. Rev. D {\bf 84}, 116003 (2011).

\bibitem{Bishara:2013vya}
F.~Bishara {\em et~al.},
\newblock JHEP {\bf 1404}, 084 (2014).

\bibitem{Harnik:2013aja}
R.~Harnik, A.~Martin, T.~Okui, R.~Primulando, and F.~Yu,
\newblock Phys. Rev. D {\bf 88}, 076009 (2013).

\bibitem{Berge:2013jra}
S.~Berge, W.~Bernreuther, and H.~Spiesberger,
\newblock Phys. Lett. B {\bf 727}, 488 (2013).

\bibitem{Modak:2013sb}
A.~Menon, T.~Modak, D.~Sahoo, R.~Sinha, and H.~Cheng,
\newblock Phys. Rev. D {\bf 89}, 095021 (2014).

\bibitem{Chen:2014gka}
Y.~Chen, R.~Harnik, and R.~Vega-Morales,
\newblock Phys. Rev. Lett. {\bf 113}, 191801 (2014).

\bibitem{Gainer:2013rxa}
J.~S. Gainer, J.~Lykken, K.~T. Matchev, S.~Mrenna, and M.~Park,
\newblock Phys. Rev. Lett. {\bf 111}, 041801 (2013).

\bibitem{Avery:2012um}
P.~Avery {\em et~al.},
\newblock Phys. Rev. D {\bf 87}, 055006 (2013).

\bibitem{Gainer:2014hha}
J.~S. Gainer, J.~Lykken, K.~T. Matchev, S.~Mrenna, and M.~Park,
\newblock Phys. Rev. D {\bf 91}, 035011 (2014).

\bibitem{Chen:2013waa}
M.~Chen {\em et~al.},
\newblock Phys. Rev. D {\bf 89}, 034002 (2014).

\bibitem{Stolarski:2012ps}
D.~Stolarski and R.~Vega-Morales,
\newblock Phys. Rev. D {\bf 86}, 117504 (2012).

\bibitem{Buchmuller:1985jz}
W.~Buchmuller and D.~Wyler,
\newblock Nucl .Phys. B {\bf 268}, 621 (1986).

\bibitem{Grzadkowski:2010es}
B.~Grzadkowski, M.~Iskrzynski, M.~Misiak, and J.~Rosiek,
\newblock JHEP {\bf 1010}, 085 (2010).

\bibitem{JHU2}
S.~Bolognesi {\em et~al.},
\newblock {Phys. Rev. D} {\bf {86}}, {095031} ({2012}).

\bibitem{JHU3}
I.~Anderson {\em et~al.},
\newblock Phys. Rev. D {\bf 89}, 035007 (2014).

\bibitem{asymmetries}
R.~M. Godbole, D.~J. Miller, and M.~M. Muhlleitner,
\newblock JHEP {\bf 0712}, 031 ({2007}).

\bibitem{Alwall:2011uj}
J.~Alwall, M.~Herquet, F.~Maltoni, O.~Mattelaer, and T.~Stelzer,
\newblock JHEP {\bf 1106}, 128 (2011).

\bibitem{Artoisenet:2013puc}
P.~Artoisenet {\em et~al.},
\newblock JHEP {\bf 1311}, 043 (2013).

\bibitem{JHU_manual}
I.~Anderson {\em et~al.},
\newblock 2014, http://www.pha.jhu.edu/spin/- -manJHUGenerator.v4.8.1.pdf .

\bibitem{pythia}
T.~Sjostrand, S.~Mrenna, and P.~Z. Skands,
\newblock {JHEP} {\bf {05}}, {026} ({2006}).

\bibitem{ATLAS_paper}
{ATLAS Collaboration},
\newblock JINST {\bf 3}, S08003 (2008).

\bibitem{CMS_paper}
{CMS Collaboration},
\newblock JINST {\bf 3}, S08004 (2008).

\bibitem{ATLAS-Collaboration:2013jwa}
{ATLAS Collaboration},
\newblock Report No. ATL-PHYS-PUB-2013-013, 2013,
  http://cds.cern.ch/record/1611123 .

\bibitem{Buchalla:2013mpa}
G.~Buchalla, O.~Cata, and G.~D'Ambrosio,
\newblock Eur .Phys. J. C {\bf 74}, 2798 (2014).

\end{thebibliography}

\end{document}